%% file: main.tex
\documentclass[twocolumn]{autart}

\usepackage{graphicx}
                       
\usepackage{amsmath}   
\usepackage{amssymb}  
                  
\usepackage[T1]{fontenc}    
\usepackage{url}            
\usepackage{amsfonts}       
\usepackage{color}         
\usepackage{cite}
\usepackage{multirow}

\newtheorem{assumption}{Assumption}
\newtheorem{remark}{Remark}
\newtheorem{theorem}{Theorem}
\newtheorem{lemma}{Lemma}
\newtheorem{definition}{Definition}
\newtheorem{proposition}{Proposition}

\usepackage{tikz}
\usetikzlibrary{patterns.meta}
\usetikzlibrary{arrows.meta}
\usetikzlibrary{plotmarks}
\usepackage{pgfplots}
\pgfplotsset{compat=newest}
\newlength\hohe
\newlength\breite
\usepackage{tabularx}

\newcommand{\regret}{\mathcal R_T}
\newcommand{\norm}[1]{\left\lVert#1\right\rVert}
\newcommand{\refeq}[1]{\overset{#1}{=}}
\newcommand{\refleq}[1]{\overset{#1}{\leq}}

\newcommand{\overbar}[1]{\mkern 1.5mu\overline{\mkern-1.5mu#1\mkern-1.5mu}\mkern 1.5mu}
\newcommand{\MAS}{\mathcal{O}}
\newcommand{\stst}{\mathcal{S}}
\newcommand{\intSv}{\overbar{\stst}_\mathrm{v}}
\newcommand{\diamS}{d_{\intSv}}
\newcommand{\constraintset}{\mathcal{Z}}
\newcommand{\ball}[1][]{\mathbb{B}^{#1}}
\newcommand{\ststcost}{L^\mathrm{s}_t}

\begin{document}

\begin{frontmatter}

\title{Online convex optimization for constrained control of nonlinear systems\thanksref{footnoteinfo}} 

\thanks[footnoteinfo]{Corresponding author M.~Nonhoff. Tel. +49 511 762 4518.\\ This work was was supported by the Deutsche Forschungsgemeinschaft (DFG, German Research Foundation) - 505182457.\\ Johannes K\"ohler was supported as a part of NCCR Automation, a National Center of Competence in Research, funded by the Swiss National Science Foundation (grant number 51NF40\_225155).}

\author[Nonhoff]{Marko Nonhoff}\ead{nonhoff@irt.uni-hannover.de},    
\author[Koehler]{Johannes K\"ohler}\ead{jkoehle@ethz.ch},               
\author[Nonhoff]{Matthias A. M\"uller}\ead{mueller@irt.uni-hannover.de}  

\address[Nonhoff]{Leibniz University Hannover, Institute of Automatic Control}  
\address[Koehler]{ETH Z\"urich, Institute for Dynamic Systems and Control}             


\begin{abstract}                          
This paper proposes a modular approach that combines the online convex optimization framework and reference governors to solve a constrained control problem featuring time-varying and a priori unknown cost functions. Compared to existing results, the proposed framework is uniquely applicable to nonlinear dynamical systems subject to state and input constraints. Furthermore, our method is general in the sense that we do not limit our analysis to a specific choice of online convex optimization algorithm or reference governor. We show that the dynamic regret of the proposed framework is bounded linearly in both the dynamic regret and the path length of the chosen online convex optimization algorithm, even though the online convex optimization algorithm does not account for the underlying dynamics. We prove that a linear bound with respect to the online convex optimization algorithm's dynamic regret is optimal, i.e., cannot be improved upon. Furthermore, for a standard class of online convex optimization algorithms, our proposed framework attains a bound on its dynamic regret that is linear only in the variation of the cost functions, which is known to be an optimal bound. Finally, we demonstrate implementation and flexibility of the proposed framework by comparing different combinations of online convex optimization algorithms and reference governors to control a nonlinear chemical reactor in a numerical experiment.
\end{abstract}

\end{frontmatter}

\section{Introduction}

The online convex optimization (OCO) framework \cite{Hazan16,Shalev12}, an extension of classical numerical optimization for online learning, has become increasingly popular over the recent years. In OCO, the goal is to minimize a sequence of cost functions that are sequentially revealed to the learner. More specifically, an OCO algorithm has to first choose an input at each time instance solely based on previous cost functions, but without information about the current one. Only then, after the input is applied, the current cost function is revealed to the algorithm, and it incurs a corresponding cost. This framework has been successfully implemented in a wide range of applications, e.g., in power systems \cite{Tang17}, trajectory planning \cite{Zheng20}, and temperature control~\cite{Zhang16}. However, frequently the presence of an underlying dynamical system hinders direct application of the OCO framework. In particular, in many applications the cost functions do not only depend on the chosen input of the OCO algorithm, but also on the physical states of an underlying dynamical system. These states cannot be arbitrarily assigned by the OCO algorithm, but result from the system dynamics. Furthermore, state and input constraint are ubiquitous in practice due to, e.g., actuator limitations, mechanical restrictions, and safety considerations, and satisfaction of these constraints at each time instance is of paramount importance. In order to tackle these challenges, we combine OCO with tools from control theory to develop a framework that minimizes time-varying and a priori unknown cost functions as described above, while handling nonlinear dynamical systems subject to state and input constraints.

Combinations of the OCO framework with control theory to solve various constrained optimal control problems have received significant interest recently. Most prominently, the so-called nonstochastic control framework \cite{Hazan22} has been studied extensively. Originating in \cite{Agarwal19}, this framework generally considers a perturbed linear system and aims to optimize a linear feedback policy using the OCO framework. In particular, in \cite{Li21a,Zhou23b} it is shown that satisfaction of state and input constraints can be guaranteed within the nonstochastic control framework by suitably defining a set of safe policies and applying an OCO algorithm that chooses a policy from the safe set at each iteration. Sublinear regret bounds with respect to the best fixed linear control policy in hindsight can be derived \cite{Foster20} by reducing the control problem to an OCO problem with memory (OCO-M), compare, e.g., \cite{Anava15}. However, this choice of benchmark in the regret definition implies that this framework studies the problem of disturbance rejection, i.e., the system is held in a vicinity of a fixed steady state and the adversarial perturbations are rejected in an optimal manner. In contrast, we consider problems where stabilizing a fixed steady state is not desirable, but the system should follow a sequence of optimal steady states, which depend on the a priori unknown and time-varying cost functions.

Furthermore, in the so-called feedback optimization framework \cite{Simonetto20}, a dynamical system is steered to the optimal solution of a (potentially time-varying) optimization problem, i.e., the optimal steady state. To this end, the control input to the system is designed by applying an optimization algorithm. In this setting, nonlinear dynamical systems are frequently considered \cite{Colombino20,Hauswirth21,Cothren22b,Chen23,He23}, but only asymptotic results in the form of stability guarantees are studied. Furthermore, constraints on the physical state of the system are generally only guaranteed to be satisfied asymptotically.

Finally, closest to this work, a number of OCO-based control approaches that aim to optimize performance of the closed loop by tracking the optimal steady state of the underlying dynamical systems have been proposed recently \cite{Karapetyan23b,Nonhoff24,Zhou23a}. In these works, dynamic regret with respect to the sequence of optimal steady states or the optimal trajectory in hindsight is typically studied. These works are able to enforce constraint satisfaction for the closed loop, but are either limited to (perturbed) linear systems and a specific choice of optimization algorithm, or the derived regret guarantees are suboptimal.

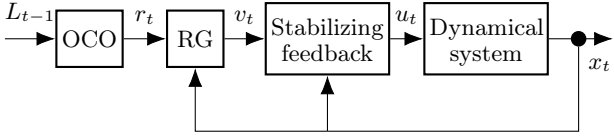
\begin{figure}
	\begin{center}
		\begin{tikzpicture}[node distance = 60,scale = 1]
			\tikzstyle{block} = [draw, minimum size=.75cm, thick, inner sep = 2pt,anchor=mid, text centered]
			\tikzstyle{add} = [draw, shape = circle, inner sep = 0pt, minimum size=.35cm, thick]
			\tikzstyle{circle} = [draw, shape=circle, inner sep = 0pt, minimum size = .15cm, fill=black, ultra thick]
			
			
			\coordinate (start) at (0,0);
			\node[block,font=\small] (OCO) [right of = start, xshift = -28] {OCO};
			\node[block,font=\small] (RG) [right of = OCO,xshift=-20] {RG};
			\node[block,text width = 1.5cm,font=\small] (contr) [right of = RG,xshift=-10] {Stabilizing feedback\strut};
			\node[block, text width = 1.5cm,font=\small] (system) [right of = contr,xshift = 0] {Dynamical system\strut};
			\node[circle] (circ) [right of = system,xshift=-25] {};			
			\coordinate[right of = circ, xshift = -47] (end);
			\coordinate[below of = contr, yshift = 25] (aux);
			
			\draw[-{Latex[length=2.5mm]}] (start) -- (OCO) node[above,midway,font=\footnotesize,yshift=.5ex] {$L_{t-1}$};
			\draw[-{Latex[length=2.5mm]}] (OCO) -- (RG) node[above,midway,font=\footnotesize,yshift=.5ex] {$r_t$};
			\draw[-{Latex[length=2.5mm]}] (RG) -- (contr) node[above,midway,font=\footnotesize,yshift=.5ex] {$v_t$};
			\draw[-{Latex[length=2.5mm]}] (contr) -- (system) node[above,midway,font=\footnotesize,yshift=.5ex] {$u_t$};
			\draw[-{Latex[length=2.5mm]}] (system) -- (end) node[below,xshift=-5,font=\footnotesize,yshift=-4] {$x_t$};
			\draw[-{Latex[length=2.5mm]}] (aux) -- (contr);
			\draw (circ) |- (aux);	
			\draw[-{Latex[length=2.5mm]}] (aux) -| (RG);
		\end{tikzpicture}
	\end{center}
	\caption{Block diagram of the proposed approach. The OCO algorithm computes a reference $r_t$ based on previous cost functions up to $L_{t-1}$, which is subsequently modified by the RG to ensure constraint satisfaction. The reference input $v_t$ from the RG is then fed to the stabilizing feedback to obtain a control input $u_t$ that is applied to the dynamical system. At time $t=0$, when no cost function is available, the OCO algorithm applies the initial reference $r_0$.}
	\label{fig:block_diagram}
\end{figure}

To summarize, prior work on control of dynamical systems using the OCO framework is typically limited to linear system dynamics, and only few algorithms are able to guarantee satisfaction of state and input constraints at each time instance. Additionally, most prior works restrict their analysis to one specific choice of OCO algorithm, most commonly online gradient descent (OGD) \cite{Zinkevich03}. In this work, we develop a modular framework that addresses these shortcomings. More specifically, the main contribution of this work is a framework that (i) is able to handle a priori unknown and time-varying cost functions as in the OCO framework, (ii) can be applied to nonlinear dynamical systems, and (iii) satisfies state and input constraints at each time instance to guarantee safe operation. Furthermore, the proposed framework is modular, i.e., does not rely on specific choices for the OCO algorithm and its components, thereby enhancing its applicability. To do so, we separate the problem of minimizing the time-varying cost functions and that of constrained control of a nonlinear dynamical system by combining a general OCO algorithm with reference governors (RGs)~\cite{Garone17}. The resulting framework is conceptually and computationally simple, and allows flexible design choices due to its modularity. The proposed framework consists of three components:
\begin{enumerate}
	\item an OCO algorithm that computes a reference based on the time-varying and a priori unknown cost functions,
	\item an RG that modifies the reference if necessary to ensure constraint satisfaction at all times, and
	\item a controller that tracks this modified reference.
\end{enumerate}
A block diagram of the proposed approach is given in Figure~\ref{fig:block_diagram}. In particular, the proposed OCO-RG framework enables a fully modular design of the OCO algorithm and the RG. More specifically, we do not limit our analysis to specific OCO algorithms and RGs. Instead, we specify important properties of the OCO algorithm and the RG for our analysis in Assumptions~\ref{ass:MAS}--\ref{ass:general_RG}, and Definition~\ref{ass:OCO-guarantees}. Crucially, we show that these properties are satisfied by almost all OCO algorithms and RGs proposed in the literature. Thus, our analysis is applicable to general OCO algorithms and RGs, and we only require Lipschitz continuity of the cost functions (but not convexity, compare Assumption~\ref{ass:cost_fcn} and Remark~\ref{rem:nonconvex_cost}), availability of a stabilizing feedback (Assumption~\ref{ass:stability}) and a steady-state mapping (Assumption~\ref{ass:steady-state-map}), and regularity of the closed-loop dynamics (Assumption~\ref{ass:Lipschitz_system}). Then, we prove that the dynamic regret of the proposed OCO-RG framework is linearly bounded by the dynamic regret and the path length of the underlying OCO algorithm (cf. Theorem~\ref{thm}). In addition, we show that such a linear dependence on the OCO algorithm's dynamic regret is optimal (cf. Proposition~\ref{prop:optimal_lower_bound}), i.e., cannot be improved upon. Furthermore, we show for $Q$-linear convergent OCO algorithms, that the dynamic regret of the proposed OCO-RG framework is bounded linearly by the variation of the cost functions only (cf. Corollary~\ref{cor:regret_exp_conv_OCO_algo}). Hence, in this case, our regret bound matches the optimal bound in~\cite{Li19} for linear systems, despite the underlying constrained nonlinear dynamical system. Moreover, we discuss that the OCO with memory (OCO-M) framework \cite{Anava15,Li18,Zhao24} is included in our setting as a special case (cf. Remark~\ref{rem:OCO-M}), which implies that the results discussed above also hold for OCO-M. Finally, we demonstrate implementation and compare different OCO algorithms and RGs within the proposed OCO-RG framework on a numerical experiment of a nonlinear chemical reactor, all of which are covered by the theory developed in this work.

We close this section by noting the preliminary results in our conference paper~\cite{Nonhoff23b}, where we first proposed to combine an OCO algorithm with an RG for linear systems. We improve the results presented therein significantly in multiple directions. First, we consider nonlinear dynamical systems subject to nonlinear state and input constraints. In contrast, the algorithm design in~\cite{Nonhoff23b} leverages a scalar RG based on a contractive maximal admissible set combined with online gradient descent, which heavily relies on linearity of the system dynamics and is generally not applicable to nonlinear systems. Furthermore, we generalize the proposed algorithm in~\cite{Nonhoff23b} to obtain a fully modular framework that features general RGs and OCO algorithms as discussed above, thereby relaxing the restrictive requirement of contractivity.

\textbf{Notation.} The set of natural numbers (including $0$) is denoted by $\mathbb{N}$, and $\mathbb{R}$ is the set of real numbers. The set of natural numbers in the interval $[a,b]$ ($[b,\infty)$) for $a>b\geq0$ is $\mathbb{N}_{[a,b]}$ ($\mathbb{N}_{\geq b}$). A symmetric positive definite matrix $A\in\mathbb{R}^{n\times n}$ is denoted by $A\succ0$. For vectors $x\in\mathbb{R}^n$, $y\in\mathbb{R}^m$, $\norm{x}$ is the Euclidean norm, we define $\norm{x}_R^2:=x^\top Rx$ for any $R\succ0$, and $(x,y):=\begin{bmatrix} x^\top & y^\top \end{bmatrix}^\top$. For a matrix $A\in\mathbb{R}^{n\times m}$, $\norm{A}$ is the induced matrix $2$-norm. A ball with dimension $n$, radius $r$, and center $x\in\mathbb{R}^n$ is given by $\ball[n]_r(x)\subseteq\mathbb{R}^n$. A continuous function $\rho:\mathbb{R}_{\geq0}\mapsto\mathbb{R}_{\geq0}$ is of class $\mathcal{K}$ if it is strictly increasing and $\rho(0) = 0$.

\textbf{Organization.} The research problem we consider is defined in Section~\ref{sec:setting}. In Section~\ref{sec:OCO-RG}, we propose our framework and discuss considered assumptions on the stabilizing feedback, the RG, and the OCO algorithm. A bound on the dynamic regret and constraint satisfaction are proven in Section~\ref{sec:theory}. Implementation of the proposed framework is demonstrated on a numerical experiment of a chemical reactor in Section~\ref{sec:example}. Section~\ref{sec:conclusion} concludes the paper. All proofs are deferred to the appendix.

\section{Problem setup} \label{sec:setting}

In this work, we consider nonlinear dynamical systems
\begin{equation}
	x_{t+1} = f(x_t,u_t),\quad t\in\mathbb{N}, \label{eq:original_system}
\end{equation}
with some initial state $x_0\in\mathcal{X}_0$, where $\mathcal{X}_0\subseteq\mathbb{R}^n$ is compact, $x_t\in\mathbb{R}^n$ denotes the state of the system and $u_t\in\mathbb{R}^m$ is the control input, both at time $t\in\mathbb{N}$. We assume that the system dynamics $f:\mathbb{R}^n\times \mathbb{R}^m\mapsto\mathbb{R}^n$ are known. System~\eqref{eq:original_system} is subject to general constraints
\begin{equation}
	(x_t,u_t) \in \constraintset, \quad t\in\mathbb{N}, \label{eq:original_constraints}
\end{equation}
i.e., the system state and control input $(x_t,u_t)$ have to be confined to the constraint set $\constraintset\subseteq\mathbb{R}^n\times\mathbb{R}^m$ at all times $t\in\mathbb{N}$. Note that both, pure input and state constraints as well as mixed constraints, can be considered by a suitable choice of the constraint set $\constraintset$. Our goal is to optimize closed-loop performance measured by time-varying and a priori unknown cost functions $L_t:\mathbb{R}^n\times\mathbb{R}^m\mapsto\mathbb{R}$, $t\in\mathbb{N}$, i.e., we aim to solve the optimal control problem
\begin{equation}
	\min_{u_0,\dots,u_T} \sum_{t=0}^T L_t(x_t,u_t) \quad \text{s.t. } \eqref{eq:original_system},~\eqref{eq:original_constraints} \label{eq:OCP}
\end{equation}
where $T\in\mathbb{N}$, and the cost functions are revealed sequentially. More specifically, at each time $t\in\mathbb{N}$, we have to commit to a control input $u_t$ based solely on previous cost functions $L_0,\dots,L_{t-1}$ and the measured system state $x_t$. Only after the control input $u_t$ is applied to system~\eqref{eq:original_system}, the current cost function $L_t$ is revealed, which leads to a cost $L_t(x_t,u_t)$. As discussed above, this setting frequently arises in a wide range of applications, such as tracking time-varying references \cite{Zheng20} or optimization of economic objectives with time-varying (energy) prices \cite{Tang17}.

\section{The proposed OCO-RG framework} \label{sec:OCO-RG}

In this section, we develop our framework to address the optimal control problem~\eqref{eq:OCP}. Note that obtaining the exact solution of~\eqref{eq:OCP} is generally intractable due to the a priori unknown and time-varying cost functions. Therefore, a common strategy in this line of research \cite{Simonetto20} is to approximate the optimal performance by tracking the (time-varying and a priori unknown) optimal steady states of system~\eqref{eq:original_system}. To this end, we first stabilize the system by designing a feedback controller. Then, we design an RG to ensure closed-loop constraint satisfaction. Finally, we use an OCO algorithm to handle the time-varying and a priori unknown cost functions. To do so, we set up a time-varying optimization problem and apply the OCO algorithm to the most recently revealed cost function in order to iteratively track the optimal solution. The output of the OCO algorithm $r_t$ is fed to the RG as the reference signal. This modular approach of combining an OCO algorithm and an RG is conceptually simple and allows us to reduce \eqref{eq:OCP} to a standard OCO problem without an underlying dynamical system. Furthermore, it enables us to derive a general framework, instead of limiting our analysis to specific choices for the RG and OCO algorithm in the proposed OCO-RG framework. In order to capture general components in our analysis, we specify required properties of the stabilizing feedback (Section~\ref{subsec:stabilizing_feedback}), the RG (Section~\ref{subsec:RG}), and the OCO algorithm (Section~\ref{subsec:OCO}) next. Crucially, these properties are satisfied by almost all RGs and OCO algorithms in the literature, and there exist constructive approaches to satisfy these assumptions by design of the feedback, the RG, and the OCO algorithm.

\subsection{Stabilizing feedback} \label{subsec:stabilizing_feedback}

The first component of our proposed OCO-RG framework is a stabilizing controller, which tracks the reference input $v_t\in\mathbb{R}^o$ provided by the RG. In order to formulate the closed-loop dynamics emerging from application of the stabilizing controller $g:\mathbb{R}^n\times\mathbb{R}^o\mapsto\mathbb{R}^m$, we first parameterize the steady states of system~\eqref{eq:original_system} using the reference input $v\in\mathbb{R}^o$, and assume that we have access to a steady-state mapping as follows.
\begin{assumption} \label{ass:steady-state-map}
	There exist a mapping $h:\mathbb{R}^o\mapsto\mathbb{R}^n$ that satisfies
	\[
		h(v)= f\Big(h(v),g\big(h(v),v\big)\Big)
	\]
	for all $v\in\mathbb{R}^o$.
\end{assumption}
In Assumption~\ref{ass:steady-state-map}, we use a reference input $v\in\mathbb{R}^o$ to parameterize the steady states of system~\eqref{eq:original_system}. Most commonly, the reference input $v$ is used as a shift in the control input, i.e., $g(x,v)=v+K(x-h(v))$. For the special case of stabilizable linear systems $f(x,u)=Ax+Bu$, Assumption~\ref{ass:steady-state-map} is trivially satisfied with a linear feedback $g(x,v)=v+Kx$, and a linear steady-state mapping $h(v)=(I_n-A_K)^{-1}Bv$. Furthermore, existence of a steady-state mapping $h$ is a standard assumption in the feedback optimization literature, compare, e.g., \cite{Hauswirth21,Chen23}, and Assumption~\ref{ass:steady-state-map} is satisfied, if the closed loop emerging from application of the feedback $g(x,v)$ is stable and continuously differentiable, compare \cite[Remark~1]{Limon18},\cite{Berberich22}. For simplicity, we assume that the steady-state mapping $h$ is known. This information is used in the OCO algorithm, which, however, can also be implemented without explicit access to the mapping $h$ (cf. Section~\ref{subsec:OCO}). In the remainder of this work, we define the control input
\begin{equation}
	u_t := g(x_t,v_t) \label{eq:control_input}
\end{equation}
for all $t\in\mathbb{N}$. Next, we apply this control input to reformulate the system dynamics~\eqref{eq:original_system} and constraints~\eqref{eq:original_constraints}, which yields the closed-loop dynamics and constraints
\begin{subequations} \label{eq:sys}
	\begin{align}
		x_{t+1} &= f_g(x_t,v_t) := f\big(x_t,g(x_t,v_t)\big), \label{eq:dynamics} \\		  
		(x_t,v_t) &\in \constraintset_g, \label{eq:constraints}
	\end{align}
\end{subequations}
where $\constraintset_g:=\big\{(x,v)\in\mathbb{R}^n\times\mathbb{R}^o~|~\big(x,g(x,v)\big)\in\constraintset\big\}$. Furthermore, we denote by $\Phi(\chi,\nu,t)$ the state $x_t$ at time $t\in\mathbb{N}$ when evolving according to~\eqref{eq:dynamics} with initial condition $\Phi(\chi,\nu,0)=\chi$ and a constant reference input $v_t\equiv\nu$. Additionally, for system~\eqref{eq:sys} we define
\begin{itemize}
	\item the set of admissible steady states 
	\[	
		\stst := \big\{(x,v)\in\mathbb{R}^n\times\mathbb{R}^o~|~x=f_g(x,v),~(x,v) \in \constraintset_g \big\},
	\]
	\item the set of admissible reference inputs
	\[	
		\stst_\mathrm{v}:= \big\{ v\in\mathbb R^o~|~\exists x\in\mathbb R^n: (x,v)\in\stst \big\},
	\]
	\item and the set of admissible system states
	\[	
		\mathcal{X}:= \big\{ x\in\mathbb{R}^n~|~\exists v\in\stst_\mathrm{v}:(x,v)\in\constraintset_g \big\}.
	\]
\end{itemize}
Furthermore, we require some regularity of the above sets and system dynamics as follows.
\begin{assumption} \label{ass:Lipschitz_system}
	The set $\stst_\mathrm{v}$ is compact. Furthermore, there exist Lipschitz constants $l_f,l_g,l_h>0$ such that
	\begin{align*}
		\norm{f_g(x,v)-f_g(\tilde{x},\tilde{v})} &\leq l_f\norm{(x,v)-(\tilde{x},\tilde{v})}, \\
		\norm{g(x,v)-g(\tilde{x},\tilde{v})} &\leq l_g\norm{(x,v)-(\tilde{x},\tilde{v})}, \\
		\norm{h(v)-h(\tilde{v})} &\leq l_h \norm{v-\tilde{v}}
	\end{align*}
	hold for all $x,\tilde{x}\in\mathcal{X}$ and $v,\tilde{v}\in\stst_\mathrm{v}$.
\end{assumption}
Finally, the following assumption formalizes the considered stability properties of the feedback $g(x,v)$.
\begin{assumption} \label{ass:stability}
	There exist constants $\lambda\in[0,1)$ and $c_\Phi\geq 1$ such that
	\begin{equation}
		\norm{\Phi(x,v,t)-h(v)}\leq c_\Phi\norm{x-h(v)}\lambda^t \label{eq:stability_bound}
	\end{equation}
	holds for any $x\in\mathcal{X}$, $v\in\stst_\mathrm{v}$, and all $t\in\mathbb N$.
\end{assumption}
Exponential stability of the closed-loop system as in Assumption~\ref{ass:stability} is commonly assumed in the related literature \cite{Agarwal19,Cothren22b,He23} and methods to design such stabilizing controllers for different classes of nonlinear systems are well studied~\cite{Khalil02}. Hence, constructive methods exist to satisfy Assumption~\ref{ass:stability} for many practical applications.

\subsection{Reference governor} \label{subsec:RG}

In this section, we describe the RG in the proposed OCO-RG framework. The RG guarantees constraint satisfaction $(x_t,v_t)\in\constraintset_g$, and, thus, $(x_t,u_t)\in\constraintset$, for all times $t\in\mathbb{N}$ by overwriting the desired reference $r_t$ with an admissible reference input $v_t\in\intSv$ based on the current system state $x_t$, where $\intSv\subseteq\stst_\mathrm{v}$ is a compact subset of $\stst_\mathrm{v}$. Even if $r_t\in\intSv$, which ensures that the desired reference satisfies the constraints \eqref{eq:constraints} at steady state, i.e., $(h(r_t),r_t)\in\constraintset_g$, the RG is necessary to guarantee constraint satisfaction during the transient phase, i.e., when $x_t\neq h(r_t)$. More specifically, RGs are generally based on constructing a forward invariant safe set $\MAS\subseteq\mathcal{X}\times\intSv$ for the closed-loop dynamics~\eqref{eq:dynamics} with constant reference input $v_t\equiv v\in\intSv$. Then, the RG chooses the reference input $v_t$ such that $(x_t,v_t)\in\MAS$. Furthermore, recall that $\Phi(\chi,\nu,t)$ denotes the evolution of the states $x_t$ of the stabilized system~\eqref{eq:dynamics} for a constant reference input $v_t\equiv\nu$. We require that the safe set $\MAS$ is constructed to have the following properties.
\begin{assumption} \label{ass:MAS}
	For all $(x,v)\in\MAS$, it holds that $\big(\Phi(x,v,t),v\big)\in\constraintset_g$ is satisfied for all $t\in\mathbb{N}$. Furthermore, there exists $\delta > 0$ such that $\ball[n]_\delta\big(h(v)\big) \subseteq\MAS_\mathrm{x}(v)$ for all $v\in\intSv$, where $\MAS_\mathrm{x}(v) := \{x\in\mathcal{X}~|~(x,v)\in\MAS\}$ is the cross section of $\MAS$ at $v\in\intSv$.
\end{assumption}
Assumption \ref{ass:MAS} ensures that $(x_t,v_t)\in\MAS$ implies satisfaction of the constraints~\eqref{eq:constraints} for all future times given a constant reference input $v\in\intSv$. There exist different constructive methods in the literature to compute a safe set $\MAS$ that satisfies Assumption~\ref{ass:MAS}, e.g., the maximal invariant set for linear systems \cite{Gilbert91} and specific classes of nonlinear systems \cite{Hirata08}, or based on a Lyapunov function \cite{Garone16}, compare also Section~\ref{sec:example} and Appendix~\ref{appendix:MAS} for more details. Additionally, Assumption~\ref{ass:MAS} requires that a (small) neighborhood of any steady state $h(v)$, $v\in\intSv$, lies in the safe set~$\MAS$. Thereby, we ensure that the references strictly satisfy the constraints, which is a standard assumption in constrained tracking control problems \cite[Assumption~1]{Koehler20}. Furthermore, we note that this assumption is not restrictive and can generally be satisfied by a suitable choice of $\intSv\subseteq\stst_\mathrm{v}$ for all the methods to construct the safe set $\MAS$ described above. Next, we describe the general RG $\mathcal{RG}:\mathcal{X}\times\intSv\mapsto\intSv$ that computes a reference input $v_t=\mathcal{RG}(x_t,r_t)$ based on the desired reference $r_t\in\intSv$ and system state $x_t$. To this end, we let $\MAS_\mathrm{v}(x) := \{v\in\intSv~|~(x,v)\in\MAS\}$ be the cross section of $\MAS$ at $x\in\mathcal{X}$.
\begin{assumption} \label{ass:general_RG}
	The RG $v_t = \mathcal{RG}(x_t,r_t)$ satisfies $(x_t,v_t)\in\MAS$ for all $t\in\mathbb{N}$. \\	
	Furthermore, denote by $\overbar{\nu}_t\geq0$ the largest value such that $\{v\in\intSv|\norm{v-v_{t-1}}\leq\overbar{\nu}_t\}\subseteq\MAS_\mathrm{v}(x_t)$ holds.  There exists a class $\mathcal K$-function $\rho:\mathbb{R}_{\geq0}\mapsto[0,1]$ such that, for all $t\in\mathbb{N}_{\geq1}$,
	\begin{itemize}
		\item if $r_t\in\MAS_\mathrm{v}(x_t)$, then $v_t = r_t$,
		\item if $r_t\notin\MAS_\mathrm{v}(x_t)$, it holds that $\norm{v_t-v_{t-1}} \geq \overbar{\nu}_t$ and 
		\begin{equation}
			\norm{r_t-v_{t}} \leq \big(1-\rho\left(\norm{v_t-v_{t-1}}\right)\big)\norm{r_t-v_{t-1}}. \label{eq:RG_contraction}
		\end{equation}
	\end{itemize}
\end{assumption}
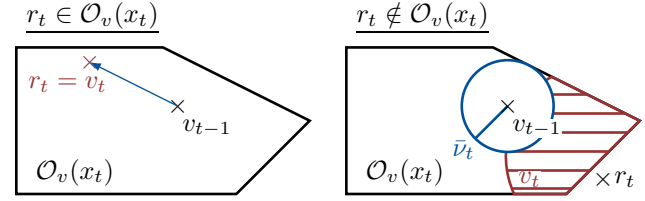
\begin{figure}
	\centering
	\definecolor{luhblue}{RGB} {0,80,155}
	\definecolor{dred}{rgb}{0.8,0.3,0.3}
	\begin{tikzpicture}[scale=.975,line width = 1pt]
		\coordinate (center1) at (2.2,1.2);
		\coordinate (center2) at (6.7,1.2);		
		\node[anchor=south west] at (0,2.05) {\underline{$r_t\in\MAS_\mathrm{v}(x_t)$}};
		\node[anchor=south west] at (4.5,2.05) {\underline{$r_t\notin\MAS_\mathrm{v}(x_t)$}};		
		\draw (0,2) -- (0,0) -- (3,0) -- (4,1) -- (2,2) -- cycle;
		\draw (4.5,2) -- (4.5,0) -- (7.5,0) -- (8.5,1) -- (6.5,2) -- cycle;
		\draw[dred,pattern color=dred,pattern={Lines[angle=-45,distance=.25cm]}] (7.3,1.6) arc [start angle=120, end angle=204.5, x radius=1.25, y radius=1.25] -- (7.5,0) -- (8.5,1) -- cycle;
		\draw[luhblue,fill = white] (center2) circle (0.626);
		\draw[luhblue] (center2.center) -- (6.257,0.757);
		\node at (center1) {$\times$};
		\node at (center2) {$\times$};
		\node[fill=white,inner sep = 0.1pt,text=luhblue] at (6.1,0.6) {$\bar\nu_t$};
		\node[fill= white, inner sep = 0.1pt] at (2.6,0.9) {$v_{t-1}$};
		\node[fill= white, inner sep = 0.1pt] at (7.1,0.9) {$v_{t-1}$};
		\node[fill= white, inner sep = 0.1pt] at (0.8,0.3) {$\MAS_\mathrm{v}(x_t)$};
		\node[fill= white, inner sep = 0.1pt] at (5.3,0.3) {$\MAS_\mathrm{v}(x_t)$};
		\node[text=dred] (rt1) at (1,1.8) {$\times$};
		\draw[luhblue,-{Latex[length=1.5mm]},line width = 0.5pt] (center1.center) -- (rt1.center);
		\node[fill= white, inner sep = 0.1pt, text=dred] at (0.7,1.5) {$r_t=v_{t}$};
		\node[fill= white, inner sep = 0.1pt,text=dred] at (7,0.2) {$v_t$};
		\node at (8,0.2) {$\times$};
		\node[fill= white, inner sep = 0.1pt] at (8.3,0.2) {$r_t$};
	\end{tikzpicture}
	\caption{Illustration of the case distinction in Assumption~\protect{\ref{ass:general_RG}}. In case that $r_t\notin\MAS_\mathrm{v}(x_t)$, Assumption~\protect{\ref{ass:general_RG}} requires that $v_t$ is chosen from the red shaded area.}
	\label{fig:RG}
\end{figure}
Assumption~\ref{ass:general_RG} is illustrated in Figure~\ref{fig:RG}. This condition ensures that $\overbar{\nu}_t>0$ whenever $v_{t-1}$ is strictly inside $\MAS_\mathrm{v}(x_t)$. In our theoretical analysis below, we show that $\overbar{\nu}_t>0$ holds eventually over any sufficiently long horizon. Together with~\eqref{eq:RG_contraction}, this implies a sufficient rate of convergence of the reference input $v_t$ to the desired reference $r_t$. The conditions in Assumption~\ref{ass:general_RG} are satisfied for standard choices of $\MAS$ by almost all RG designs in the literature \cite{Garone17}, such as the scalar RG which chooses 
\begin{subequations} \label{eq:scalar_RG}
\begin{align}
	\beta_t &=
	\begin{cases}
	 	&\arg\max_{\beta\in[0,1]} \beta \\
		&\text{s.t. } \big(x_t,v_{t-1}+\beta(r_t-v_{t-1})\big)\in\MAS, 
	\end{cases} \\
	v_t &= v_{t-1}+\beta_t(r_t-v_{t-1})
\end{align}
\end{subequations}
and the command governor that sets 
\[
	v_t=\arg\min_{v\in\intSv}\norm{v-r_t}^2 \quad \text{s.t. } (x_t,v)\in\MAS. 
\]
In more detail, most RGs set $v_t=r_t$ if possible, and choose the reference input $v_t$ as close as possible to the desired reference $r_t$ if $r_t\notin\MAS_\mathrm{v}(x_t)$, thereby satisfying Assumption~\ref{ass:general_RG}. In contrast, we only require to move the reference input $v_t$ towards the desired reference $r_t$ by a minimum amount $\overbar{\nu}_t$ in~\eqref{eq:RG_contraction}. To the best of the authors' knowledge, a general characterization of RGs as in Assumption~\ref{ass:general_RG} was previously not available in the literature. Note that, for $t=0$, Assumption~\ref{ass:general_RG} ensures that the initialization $v_0$ of the RG is feasible, i.e., $(x_0,v_0)\in\MAS$, and that $(x_t,v_t)\in\MAS$ for all $t\in\mathbb{N}$, which implies $(x_t,v_t)\in\constraintset_g$ for all $t\in\mathbb{N}$ by Assumption~\ref{ass:MAS}.

\begin{remark} \label{rem:general_RG} (RGs without invariance and/or with weighting)
While Assumptions~\ref{ass:general_RG} is satisfied by nearly all RGs in the literature (compare the survey \cite{Garone17} and the references therein), there are some RGs which do not satisfy (parts of) this assumption. In particular, forward invariance of $\MAS$ is sometimes relaxed to `strong returnability' \cite{Gilbert02}, which means that $(x_t,v_t)\in\MAS$ may not be satisfied at all times. Typically, these RGs set $v_t=v_{t-1}$ whenever $\MAS_\mathrm{v}(x_t)=\emptyset$, thereby ensuring $(x_t,v_t)\in\constraintset_g$, compare Assumption~\ref{ass:MAS}. Our analysis can equally consider this case with minor modifications and we mainly keep the assumption $(x_t,v_t)\in\MAS$ for all $t\in\mathbb{N}$ for simplicity of exposition of our results. Second,~\eqref{eq:RG_contraction} may not hold for some RGs. For example, consider the command governor \cite{Angeli99} with weighting that chooses 
\[
	v_t=\arg\min_{v\in\intSv}\norm{v-r_t}_R^2 \quad \text{s.t. }(v,x_t)\in\MAS
\]
with some positive definite weighting matrix $R\succ0$. Again, it is straightforward to modify our analysis to include these RGs by replacing~\eqref{eq:RG_contraction} with
	\[
		\norm{r_t-v_{t}}_R \leq \big(1-\rho\left(\norm{v_t-v_{t-1}}\right)\big)\norm{r_t-v_{t-1}}_R,
	\]
	which only leads to additional constant factors in our regret analysis below.
\end{remark}

\subsection{OCO algorithm} \label{subsec:OCO}

The last component of the OCO-RG framework is an OCO algorithm that provides a desired reference $r_t$ based only on previous cost functions $L_0,\dots,L_{t-1}$. To this end, we define the steady-state cost functions $\ststcost(v):=L_t\big(h(v),g(h(v),v)\big)$, and apply an OCO algorithm to the time-varying optimization problem
\begin{equation}
	\min_{r_0,\dots,r_T\in\intSv} \sum_{t=0}^T \ststcost(r_t), \label{eq:OCO_opt_problem}
\end{equation}
where the time-varying and a priori unknown steady-state cost functions $\ststcost:\intSv\mapsto\mathbb{R}$ are revealed sequentially. Hence, the optimization problem~\eqref{eq:OCO_opt_problem} is a standard OCO problem that can be solved by a large variety of algorithms in the literature, e.g., first-order methods \cite{Jadbabaie15,Mokhtari16,Lin24}, projection-free methods \cite{Zhou23} and second-order methods \cite{Chang21,Dvurechensky24}. As discussed above, in case the steady-state mapping $h$ is not known explicitly, the OCO algorithm can be applied to the sequence of cost functions $L_t(x_t,u_t)$ by incorporating an additional equality constraint $x_t=f_g(x_t,r_t)$. In our framework, we consider a general OCO algorithm $\mathcal{A}^\mathrm{OCO}:\mathcal{I}_t\mapsto\intSv$ that computes a feasible reference $r_t\in\intSv$ based on the previous cost functions and references, i.e, $r_t=\mathcal A^\mathrm{OCO}(\mathcal I_t)$, where $\mathcal I_t := \{r_0,\dots,r_{t-1},L_0,\dots,L_{t-1}\}$ describes all available information at time $t$. For simplicity, we assume that the OCO algorithm has access to the full functional form of the previous cost functions $L_0,\dots,L_{t-1}$. However, most OCO algorithms in the literature require less information. For example, online gradient descent \cite{Zinkevich03} only evaluates $\nabla L^s_{t-1}(r_{t-1})$ at each time instance $t$. Furthermore, there exist suitable algorithms that only require a finite number of function evaluations $L_0,\dots,L_{t-1}$ (e.g., \cite{Zhao21,Yang16}), which can be implemented within the proposed OCO-RG framework. We assume that the OCO algorithm and the RG have the same initialization, i.e., $r_0=v_0\in\intSv$. This assumption simplifies notation in our analysis below and can easily be satisfied since $r_0\in\intSv$ and $v_0\in\MAS_\mathrm{v}(x_0)$ can be chosen freely. Furthermore, we define the optimal steady-state reference $\eta_t\in\intSv$ by
\begin{equation}
	\eta_t:=\arg\min_{v\in\intSv} \ststcost(v), \label{eq:def_optimal_steady_state}
\end{equation}
for all $t\in\mathbb{N}$, which implies that the optimal steady state is given by $h(\eta_t)\in\mathcal{X}$. The optimal steady-state reference $\eta_t$ is not computed by the proposed OCO-RG framework, but only serves as a benchmark in our theoretical analysis below. Then, we define the dynamic regret and path length of the OCO algorithm as follows.
\begin{definition} \label{ass:OCO-guarantees}
	The dynamic regret $\regret^\mathrm{OCO}$ and the path length $\regret^\mathrm{PL}$ of the OCO algorithm $\mathcal{A}^\mathrm{OCO}$ are defined by
	\begin{align}
		\regret^\mathrm{OCO}(r_0,\dots,r_T) &:= \sum_{t=0}^T \ststcost(r_t) - \ststcost(\eta_t), \label{eq:OCO-regret} \\
		\regret^\mathrm{PL}(r_0,\dots,r_T) &:= \sum_{t=1}^T \norm{r_t-r_{t-1}}. \label{eq:path-length-regret}
	\end{align}
\end{definition}
In the remainder of this work, we omit the arguments of $\regret^\mathrm{OCO}$ and $\regret^\mathrm{PL}$ when they are clear from context. Interestingly, the path length $\regret^\mathrm{PL}$ of the OCO algorithm $\mathcal{A}^\mathrm{OCO}$ plays a crucial role in our analysis besides its dynamic regret. Therefore, an OCO algorithm should be implemented that admits small upper bounds on $\regret^\mathrm{OCO}$ and $\regret^\mathrm{PL}$ simultaneously. Furthermore, as common in OCO-based control, we require some regularity of the cost functions.
\begin{assumption} \label{ass:cost_fcn}
	There exists a Lipschitz constant $l>0$ such that
	\begin{equation}
		\norm{L_t(x,u) - L_t(\tilde{x},\tilde{u})} \leq l \norm{(x,u)-(\tilde{x},\tilde{u})} \label{eq:Lipschitz_cost}
	\end{equation}	
	holds for all $t\in\mathbb{N}$ and $(x,u)\in\constraintset$, $(\tilde{x},\tilde{u})\in\constraintset$.
\end{assumption}

\begin{remark} \label{rem:nonconvex_cost} (Nonconvex cost functions)
	 Note that we do not assume (strong) convexity of the steady-state cost functions $L^s_t$. Thus, our proposed framework is also applicable for some classes of nonconvex cost functions $L^s_t$, compare Section~\ref{sec:example}. In particular, we show that the dynamic regret of the OCO-RG framework is bounded linearly by the dynamic regret of the OCO algorithm. Depending on the class of nonconvex cost functions, the dynamic regret of the OCO algorithm \cite{Dvurechensky24,Mulvaney-Kemp23} can increase significantly, thereby leading to a larger regret bound for our proposed framework.
\end{remark}

For a special class of OCO algorithms, it is possible to derive bounds on both their dynamic regret and path length which are linear in the path length of problem~\eqref{eq:OCO_opt_problem}. More specifically, we show next that OCO algorithms, which converge $Q$-linearly for a constant cost function, admit such bounds.

\begin{proposition} \label{prop:bounded_regret_path_length_exp_convergence}
	Suppose Assumptions~\ref{ass:steady-state-map},~\ref{ass:Lipschitz_system}, and~\ref{ass:cost_fcn} hold, and that there exist $S\succ0$ and $\kappa\in[0,1)$, such that
	\begin{equation}
		\norm{r_t-\eta_{t-1}}_S \leq \kappa \norm{r_{t-1}-\eta_{t-1}}_S \label{eq:exp_convergence}
	\end{equation}
	holds for the OCO algorithm $r_t = \mathcal{A}^\mathrm{OCO}(\mathcal{I}_t)$ for all $t\in\mathbb{N}$. Then,
	\begin{align*}
		\regret^\mathrm{OCO} &\leq \tilde{c}_\mathrm{OCO,0}\norm{r_0-\eta_0}+ \tilde{c}_\mathrm{OCO}\sum_{t=1}^T \norm{\eta_t-\eta_{t-1}}, \\
		\regret^\mathrm{PL} &\leq \tilde{c}_\mathrm{PL,0}\norm{r_0-\eta_0}+ \tilde{c}_\mathrm{PL}\sum_{t=1}^T \norm{\eta_t-\eta_{t-1}}
	\end{align*}
	hold for any sequence of cost functions $L_t$ and any $T\in\mathbb{N}$, where $\tilde{c}_{\mathrm{OCO},0}$ and $\tilde{c}_\mathrm{OCO}$ are defined in~\eqref{eq:cOCO0} and~\eqref{eq:cOCO}, respectively, and $\tilde{c}_{\mathrm{PL},0}:=\frac{1+\kappa}{l_s}\tilde{c}_{\mathrm{OCO},0}$ and $\tilde{c}_\mathrm{PL}:=\frac{1+\kappa}{l_s}\tilde{c}_\mathrm{OCO}$.
\end{proposition}
The proof is given in Appendix~\ref{appendix:Prop_bounded_regret_path_length_exp_convergence}. Optimization algorithms that achieve $Q$-linear convergence as in \eqref{eq:exp_convergence} include many first-order methods, such as projected gradient descent if the steady-state cost functions $\ststcost$ are at least weakly quasi-strongly convex, have a Lipschitz continuous gradient, and if $\intSv$ is convex, compare~\cite[Theorem~11]{Necoara19} and~\cite[Theorem~5]{Alimisis24}. Furthermore, computing the previously optimal reference $\eta_{t-1}$ and setting $r_t=\eta_{t-1}$ also satisfies~\eqref{eq:exp_convergence} with $\kappa=0$.

\subsection{The complete OCO-RG framework}

Summarizing the above, the complete OCO-RG framework is given in algorithmic form below.

\par\noindent\rule{.475\textwidth}{1.5pt} \\%
\vspace{-5pt}%
\noindent\textbf{Algorithm~1:} The OCO-RG framework%
\vspace{-10pt}%
\par\noindent\rule[0pt]{.475\textwidth}{0.6pt}
\textbf{For all $t\in\mathbb{N}$:} Given a measurement $x_t$, \\%
\begin{tabularx}{\linewidth}{rX}
	{[S1]} & compute $r_t=\mathcal{A}^\mathrm{OCO}\left(\mathcal{I}_t\right)$, \\
	{[S2]} & compute $v_t=\mathcal{RG}(x_t,r_t)$ using the safe set from Assumption~\ref{ass:MAS}, \\
	{[S3]} & apply $u_t=g(x_t,v_t)$ to system~\eqref{eq:original_system}.
\end{tabularx} \\%
\vspace{-15pt}%
\par\noindent\rule{.475\textwidth}{1.5pt}

The proposed OCO-RG framework in Algorithm~1 is fully modular, i.e., it allows to combine arbitrary OCO algorithms, RGs that satisfy Assumptions~\ref{ass:MAS} and~\ref{ass:general_RG}, and stabilizing controllers satisfying Assumptions~\ref{ass:steady-state-map}, \ref{ass:Lipschitz_system}, and~\ref{ass:stability}. These assumptions are not restrictive and satisfied by standard methods in the literature. In particular,
\begin{itemize}
	\item in [S3], we require a controller that exponentially stabilizes the dynamical system~\eqref{eq:sys}, cf. Assumption~\ref{ass:stability}. Methods to design such controllers for many practical applications are well-studied in the literature \cite{Khalil02}.
	\item safe sets $\MAS$ that are applicable in [S2], i.e., satisfy Assumption~\ref{ass:MAS}, can be computed, e.g., based on a Lyapunov function, compare~\cite{Garone16,Gilbert02,Nicotra15} and Appendix~\ref{appendix:MAS}, or based on the maximal admissible set~\cite{Hirata08,Kalabic14}. Furthermore, RGs satisfying Assumption~\ref{ass:general_RG} include almost all methods in the literature, see the survey \cite{Garone17} and the references therein.
	\item we do not pose any assumption on the OCO algorithm in [S1], compare Section~\ref{subsec:OCO}. Hence, OCO algorithms that are applicable to the constrained time-varying optimization problem~\eqref{eq:OCO_opt_problem} include first-order methods \cite{Mulvaney-Kemp23,Zinkevich03,Jadbabaie15,Mokhtari16,Lin24}, projection-free methods \cite{Zhou23}, and second-order methods \cite{Chang21,Dvurechensky24}.	
\end{itemize}

\section{Regret analysis} \label{sec:theory}

In this section, we provide the theoretical analysis of the proposed OCO-RG framework. As common in OCO, we assess the closed-loop performance of the proposed approach by its dynamic regret.
\begin{definition} \label{def:regret}
	For any $T\in\mathbb{N}$, sequence of cost functions $L_t$ and control inputs $u_t$, $t\in\mathbb{N}_{[0,T]}$, the dynamic regret is defined as
	\begin{equation}
		\regret := \sum_{t=0}^T L_{t}(x_t,u_t) - \sum_{t=0}^T \ststcost(\eta_t). \label{eq:def_regret}
	\end{equation}
\end{definition}
In \eqref{eq:def_regret}, the closed-loop cost is compared to the optimal steady-state cost at each time step. If $\regret\leq0$, then~\eqref{eq:def_regret} implies that the closed-loop performance (i.e., the accumulated closed-loop cost) would be at least as good as jumping to the optimal steady states exactly. However, note that it is generally impossible to design a sequence of control inputs $u_t$ such that system~\eqref{eq:original_system} tracks the state sequence given by $h(\eta_t)$ exactly, because the sequence $h(\eta_t)$ does in general not follow the system dynamics~\eqref{eq:original_system} and is a priori unknown. Therefore, it has been shown in the literature that the best achievable regret guarantee is lower bounded by the path length of the problem $\sum_{t=1}^T \norm{\eta_t-\eta_{t-1}}$ \cite{Li19}, and connections of such a regret formulation to asymptotic stability have been analyzed \cite{Nonhoff23a,Karapetyan23a}. Furthermore, we note that a similar performance analysis with respect to the optimal steady state is also commonly performed in, e.g., economic model predictive control~\cite{Faulwasser18} and previous works on OCO-based control \cite{Li19,Nonhoff24}. Note that the benchmark $\eta_t\in\intSv$ in~\eqref{eq:def_regret} depends on the set $\intSv$, which can be chosen freely. Typically, the set $\intSv$ should be chosen as large as possible, e.g., $\intSv=(1-\delta_s)\stst_\mathrm{v}$ with some arbitrarily small positive constant $\delta_s\in(0,1)$. In order to show our main result, we require three auxiliary results. First, we show that Assumption~\ref{ass:stability} implies existence of a suitable Lyapunov function for system~\eqref{eq:dynamics}.
\begin{lemma} \label{lem:Lyap_fcn}
	Suppose 
	Assumptions~\ref{ass:steady-state-map}--\ref{ass:stability} 
	hold. Then, there exists a function $V:\mathcal{X}\times\stst_\mathrm{v}\mapsto \mathbb R_{\geq0}$ and constants $\lambda_2\geq\lambda_1>0$, $\lambda_3>0$, and $l_V>0$, such that
	\begin{align}
		\lambda_1\norm{x-h(v)} \leq V(x,v) &\leq\lambda_2\norm{x-h(v)}, \label{eq:Lyap_fcn_bounds}\\
		V\big(f_g(x,v),v\big) - V(x,v) &\leq -\lambda_3\norm{x-h(v)} \label{eq:Lyap_fcn_decrease}
	\end{align}
	hold for all $x\in\mathcal{X}$ and $v\in\stst_\mathrm{v}$. Furthermore, 
	\begin{equation}
		\norm{V(x,v) - V(x,\tilde{v})} \leq l_V\norm{v-\tilde{v}} \label{eq:Lipschitz_Lyap_fcn}
	\end{equation}
	holds for all $x\in\mathcal{X}$ and $v,\tilde{v}\in\stst_\mathrm{v}$ that satisfy $\Phi(x,v,t)\in\mathcal{X}$ and $\Phi(x,\tilde{v},t)\in\mathcal{X}$ for all $t\in\mathbb{N}$.
\end{lemma}
The proof is given in Appendix~\ref{appendix:Lemma_Lyap_fcn}. Second, Lemma~\ref{lem:Lyap_fcn_increase_bound} provides a bound on the influence of changes of the reference input on the Lyapunov function from Lemma~\ref{lem:Lyap_fcn}.
\begin{lemma} \label{lem:Lyap_fcn_increase_bound}
	Suppose 
	Assumptions~\ref{ass:steady-state-map}--\ref{ass:general_RG} 
	hold. For any $\tau_1,\tau_2\in\mathbb{N}$ such that $\tau_2\geq\tau_1$, the Lyapunov function $V(x,v)$ from Lemma~\ref{lem:Lyap_fcn} satisfies
	\begin{equation}
		\begin{split} \label{eq:Lyap_recursion}
			V(x_{\tau_2},v_{\tau_2}) &\leq \tilde{\lambda}^{\tau_2-\tau_1} V(x_{\tau_1},v_{\tau_1}) \\
			&\qquad+ l_V\sum_{i=\tau_1+1}^{\tau_2} \norm{v_i-v_{i-1}} \tilde{\lambda}^{\tau_2-i},
		\end{split}
	\end{equation}
	where $\tilde{\lambda} := 1-\frac{\lambda_3}{\lambda_2}\in[0,1)$. Furthermore, there exists $\overbar{V}>0$ such that $V(x_t,v_t)\leq\overbar{V}$ holds for all $t\in\mathbb{N}$.
\end{lemma}
The proof is given in Appendix~\ref{appendix:Lemma_Lyap_fcn_increase_bound}. Finally, we use Lemma~\ref{lem:Lyap_fcn_increase_bound} to obtain the following result on the average convergence of the reference input $v_t$ over a horizon of $M\in\mathbb{N}$ time steps.
\begin{lemma} \label{lem:RG_avg_contr}
	Suppose 
	Assumptions~\ref{ass:steady-state-map}--\ref{ass:general_RG} 
	hold. Then, there exist $M\in\mathbb{N}$ and $\epsilon\in(0,1]$ such that
	\begin{equation}
		\prod_{i=t}^{t+M} \big(1-\rho(\alpha_i)\big) \leq 1-\epsilon \label{eq:RG_avg_contr}
	\end{equation}
	holds for all $t\in\mathbb{N}$, where
	\begin{equation}
		\alpha_t = \begin{cases} \rho^{-1}(\epsilon)&\text{if } r_t = v_t \\ \norm{v_t-v_{t-1}} &\text{otherwise} \end{cases} \label{eq:def_alpha}
	\end{equation}
	and $\rho:\mathbb R_{\geq0}\mapsto[0,1]$ is from Assumption~\ref{ass:general_RG}.
\end{lemma}
The proof is given in Appendix~\ref{appendix:Lemma_RG_avg_contr}. Next, we are ready to prove our main result, which provides an upper bound on the dynamic regret~\eqref{eq:def_regret} of the proposed framework. \pagebreak
\begin{theorem} \label{thm}
	Suppose 
	Assumptions~\ref{ass:steady-state-map}--\ref{ass:cost_fcn}
	hold. Then, the closed loop satisfies the constraints, i.e., $(x_t,u_t)\in\mathcal{Z}$ holds for all $t\in\mathbb{N}$. Furthermore, the dynamic regret of the OCO-RG framework satisfies
	\begin{equation}
		\regret \leq c_{0} + \regret^\mathrm{OCO}+c_\mathrm{PL}\regret^\mathrm{PL} \label{eq:regret_bound}
	\end{equation}
	for any initial conditions $x_0\in\MAS_\mathrm{x}(r_0)$, sequence of cost functions $L_t$, and $T\in\mathbb{N}$ with 
	\[
		c_\mathrm{PL} = l_s \frac{M}{\epsilon}+l(1+l_g)l_V\frac{2M+\epsilon}{\epsilon\lambda_1(1-\tilde{\lambda})}
	\] 
	and 
	\[
		c_0 = \frac{l(1+l_g)\lambda_2}{\lambda_1(1-\tilde{\lambda})} \big(\norm{x_0-h(\eta_0)} + l_h\norm{v_0-\eta_0}\big).
	\]
\end{theorem}
The proof of Theorem~\ref{thm} is given in Appendix~\ref{appendix:thm}. Theorem~\ref{thm} shows that the dynamic regret of the OCO-RG framework is bounded linearly in both the dynamic regret and the path length of the underlying OCO algorithm. We note that the path length of the OCO algorithm $\regret^\mathrm{PL}$ is also studied in the context of switching costs \cite{Li18,Zhao24}. Intuitively, this bound captures that frequent changes of the reference signal (leading to a large path length $\regret^\mathrm{PL}$) can excite the dynamical system \eqref{eq:dynamics}, which may lead to fluctuations, and, thus, a large closed-loop cost even if the steady-state cost remains small (leading to a small dynamic regret $\regret^\mathrm{OCO}$) due to the dependence of the cost functions $L_t(x_t,u_t)$ on the system state $x_t$. Furthermore, Theorem~\ref{thm} highlights the influence of the other two components of the OCO-RG framework on the regret bound. More specifically, the regret bound~\eqref{eq:regret_bound} decreases by choosing (i) a larger safe set for the RG, which affects the function $\rho\in\mathcal{L}$ in Assumption~\ref{ass:general_RG} and, thus, the constants $M$ and $\epsilon$, and (ii) a stabilizing controller $g$ that converges quickly, which is reflected in the Lyapunov constants $\lambda_1$, $\lambda_2$, and $\tilde{\lambda}$ from Lemmas~\ref{lem:Lyap_fcn} and~\ref{lem:Lyap_fcn_increase_bound}, as well as the constants $M$ and $\epsilon$ from Lemma~\ref{lem:RG_avg_contr}. Next, we show that the linear dependence on the dynamic regret of the OCO algorithm $\regret^\mathrm{OCO}$ is, in fact, an optimal bound.
\begin{proposition} \label{prop:optimal_lower_bound}
	For any control algorithm $u_t=\mathcal{A}(x_0,\dots,x_t,u_0,\dots,u_{t-1},L_0,\dots,L_{t-1})$ and any OCO algorithm $r_t=\mathcal{A}^\mathrm{OCO}(\mathcal{I}_t)$, there exists a sequence of strongly convex and Lipschitz continuous cost functions $L_t$ such that their dynamic regret with respect to the time-varying optimization problems~\eqref{eq:OCP} and~\eqref{eq:OCO_opt_problem}, respectively, satisfy
	\[
		\regret\big(x_0,u_0,\dots,u_T\big) \geq \regret^\mathrm{OCO}(r_0,\dots,r_T).
	\]
\end{proposition}
The proof is given in Appendix~\ref{appendix:Prop_optimal_lower_bound}. Proposition~\ref{prop:optimal_lower_bound} shows that no algorithm can achieve a bound on the dynamic regret~\eqref{eq:def_regret} for the time-varying optimal control problem~\eqref{eq:OCP} that is sublinear in $\regret^\mathrm{OCO}$ and simultaneously holds for any sequence of cost functions. Hence, the linear dependence of the upper bound~\eqref{eq:regret_bound} on the dynamic regret of the underlying OCO algorithm $\regret^\mathrm{OCO}$ is optimal and cannot be improved upon. However, an analogous lower bound for the linear dependence on the path length of the OCO algorithm $\regret^\mathrm{PL}$ does not exist in the literature. Nevertheless, for $Q$-linear convergent OCO algorithms as in Proposition~\ref{prop:bounded_regret_path_length_exp_convergence}, the regret bound~\eqref{eq:regret_bound} can be improved as follows.
\begin{cor} \label{cor:regret_exp_conv_OCO_algo}
	Suppose 
	Assumptions~\ref{ass:steady-state-map}--\ref{ass:cost_fcn}
	hold. Furthermore, suppose that the OCO algorithm $r_t=\mathcal{A}^\mathrm{OCO}(\mathcal{I}_t)$ is $Q$-linear convergent, i.e.,~\eqref{eq:exp_convergence} holds for some $S\succ0$ and $\kappa\in[0,1)$, and for any cost function $L_{t-1}$. Then,
	\begin{align}
	\begin{split}
		\regret \leq c_0+&(\tilde{c}_{\mathrm{OCO},0}+c_\mathrm{PL}\tilde{c}_\mathrm{PL,0}) \norm{r_0-\eta_0}\\ 
		&+ (\tilde{c}_\mathrm{OCO}+c_\mathrm{PL}\tilde{c}_\mathrm{PL}) \sum_{t=1}^T \norm{\eta_t-\eta_{t-1}} \label{eq:cor_bound}
	\end{split}
	\end{align}
	holds for any series of cost functions $L_t$, initial conditions $x_0\in\MAS_\mathrm{x}(r_0)$, and $T\in\mathbb{N}$.
\end{cor}
Corollary~\ref{cor:regret_exp_conv_OCO_algo} follows directly by applying~Proposition~\ref{prop:bounded_regret_path_length_exp_convergence} to the bound in Theorem~\ref{thm}. It shows that, for $Q$-linear convergent OCO algorithms, the dynamic regret of the OCO-RG framework is bounded linearly by the path length of the minimizers of~\eqref{eq:OCO_opt_problem}, which can be seen as a measure of the variation of the cost functions. It is well known that such a linear bound~\eqref{eq:cor_bound} is optimal, i.e., the best achievable bound \cite{Li19}. Furthermore, such a bound implies asymptotic stability of the optimal steady state (for a constant cost function) under mild conditions \cite{Nonhoff23a}.

\begin{remark} \label{rem:OCO-M} (Connections to OCO-M)
	The setting considered in this paper includes OCO with memory (OCO-M) \cite{Anava15} as a special case. OCO-M is a generalization of the OCO framework, where the cost function is allowed to depend on previous actions $u_{t-p},\dots,u_t\in\mathbb{R}^m$ for some $p\in\mathbb{N}$, i.e., the goal is to solve
	\begin{equation}
		\min_{\{u_t\}_{t=0}^T} \sum_{t=0}^T L_t(u_{t-p},\dots,u_t)\quad\text{s.t. }u_t\in\mathcal{U},~t\in\mathbb{N}_{[0,T]}, \label{eq:OCO-M_opt_problem}
	\end{equation}
	where the cost functions are time-varying and a priori unknown and~ $\mathcal{U}\subseteq\mathbb{R}^m$ is a constraint set. This dependence on previous actions is used most commonly to penalize large deviations between subsequent time instances by choosing $p=1$ and $L_t(u_{t-1},u_t) = \tilde{L}_t(u_t)+\norm{u_t-u_{t-1}}^2$ \cite{Li18,Zhao24}. We can reformulate~\eqref{eq:OCO-M_opt_problem} into the form~\eqref{eq:OCP} as follows: We define the states $x_t = \begin{bmatrix} u_{t-p}^\top & \dots & u_{t-1}^\top \end{bmatrix}^\top$ leading to a linear dynamical system $x_{t+1}=Ax_t+Bu_t$. It is easy to see that this system is stable and, thus, Assumptions~\ref{ass:steady-state-map}--\ref{ass:stability} are satisfied with $g(x,v)=0$, and $h(v)=Hv$ with $H=\begin{bmatrix} I_m & \dots & I_m \end{bmatrix}^\top \in \mathbb{R}^{mp\times m}$. Furthermore, the constraints~\eqref{eq:original_constraints} only affect the input, i.e, $u_t\in\mathcal{U}$. Thus, an RG satisfying Assumptions~\ref{ass:MAS} and~\ref{ass:general_RG} is simply given by $v_t=r_t$, which leads to $u_t=r_t$. Finally, we observe that the steady-state cost function $\ststcost(\nu)$ for this problem is given by $\ststcost(\nu)=L_t(\nu,\dots,\nu)$. Similar to other approaches in the OCO-M literature \cite{Li18,Zhao24}, we solve the OCO-M problem \eqref{eq:OCO-M_opt_problem} by applying an OCO algorithm to $\min_{r_0,\dots,r_T\in\mathcal{U}} \sum_{t=0}^T L_t(r_t,\dots,r_t)$ in our OCO-RG framework. Then, assuming that Assumption~\ref{ass:cost_fcn} is satisfied, Theorem~\ref{thm} yields the dynamic regret bound
	\begin{align*}
		\regret^\mathrm{OCO-M} &:= \sum_{t=0}^T L_t(u_{t-p},\dots,u_t) - \sum_{t=0}^T \min_{\nu\in\mathcal{U}} L_t(\nu,\dots,\nu) \\
		&\leq c_0 + \regret^\mathrm{OCO} + c_\mathrm{PL} \regret^\mathrm{PL}.
	\end{align*}
	Thus, our results are applicable to OCO-M and recover similar results as in, e.g.,~ \cite{Li18} as a special case.
\end{remark}

\section{Numerical example} \label{sec:example}

In this section, we demonstrate application and modularity of the proposed OCO-RG framework by comparing combinations of two different OCO algorithms and safe sets $\MAS$. The code for the simulations can be found online at \url{https://doi.org/10.25835/4igbjn07}.

\subsection{Setup}

We consider the continuous stirred tank reactor from~\cite{Mayne11}. The nonlinear dynamics~\eqref{eq:original_system} are given by an Euler forward discretization with sampling time $\tau=0.1$\,s (i.e., one discrete time step corresponds to $0.1$\,s) of the continuous-time nonlinear dynamics
\begin{equation*}
	\begin{bmatrix} \dot{c}_t \\ \dot{\vartheta}_t \end{bmatrix} = \begin{bmatrix} \frac{1}{\theta_f} (1-c_t) - kc_te^{-\frac{M}{\vartheta_t}} \\
		\frac{1}{\theta_f} (x_f-\vartheta_t) + kc_te^{-\frac{M}{\vartheta_t}}-\alpha_f u_t(\vartheta_t-x_c) \end{bmatrix}, 
\end{equation*}
with concentration $c_t$, temperature $\vartheta_t$, coolant flow rate $u_t$, and system state $x_t = \begin{bmatrix} c_t & \vartheta_t \end{bmatrix}^\top$. The parameters of the tank reactor are identical to~\cite{Mayne11}, i.e., $\theta_f =20$, $k=300$, $M=5$, $x_f=0.3947$, $x_c=0.3816$, and $\alpha_f=0.117$. The initial state is set to $x_0 =\begin{bmatrix} 0.2632 & 0.6519 \end{bmatrix}^\top$.  Note that this system does not admit a unique mapping from a constant control input $u_t$ to the steady states of the system, because the same steady-state control input corresponds to different steady states in some cases. Instead, we use the temperature $\vartheta$ as the reference input, and to parameterize the steady states of the system. Then, the resulting steady-state mapping $h(\vartheta)$ satisfies Assumption~\ref{ass:steady-state-map}, and is illustrated in Figure~\ref{fig:steady-states}.
\begin{figure}
	\centering
	\small
	\setlength\hohe{3cm}
	\setlength\breite{.4\textwidth}
	\include{figures/steadystates}
	\vspace{-5ex}
	\caption{Steady states of the tank reactor (blue line) together with the safe sets $\MAS_\mathrm{fix}$ (blue sets) and $\MAS_\mathrm{var}$ (red sets).}
	\label{fig:steady-states}
\end{figure}

The system is subject to the state and input constraints $\vartheta_t\in[0,1]$, $c_t\in[0,1]$, and $u_t\in[0,2]$, i.e., we consider the constraint set $\constraintset=[0,1]^2\times[0,2]$. Furthermore, we choose the tightened set of admissible steady-state reference inputs $\intSv$ such that $\vartheta_t\in[0.4,0.85]$ for all $v_t\in\intSv$, which corresponds to $c_t\in[0.05,0.98]$.

\begin{figure}
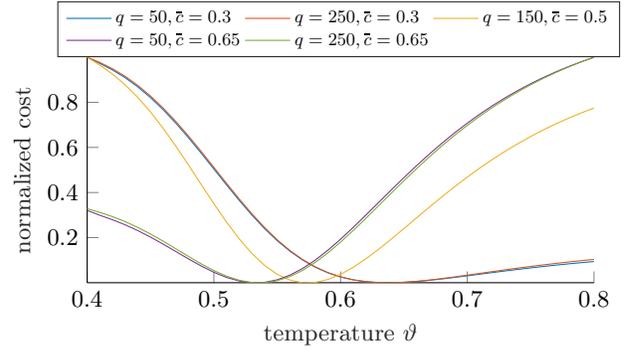

	\centering
	\small
	\setlength\hohe{3cm}
	\setlength\breite{.4\textwidth}
	\include{figures/cost}
	\vspace{-5ex}
	\caption{Depiction of the normalized steady-state cost function $L_t^s(\vartheta)$ for different values of $q$ and $\overbar{c}$.}
	\label{fig:cost}
\end{figure}

The cost functions are given by 
\[
	L_t(x,u) = q_t\norm{\begin{bmatrix} 1 & 0 \end{bmatrix} x-\overbar{c}_t}^2+u^2,
\]	
where $q_t\in[50,250]$ and $\overbar{c}_t \in [0.25,0.65]$ are a weighting parameter and the desired concentration, respectively, which are both time-varying and a priori unknown. Thus, the a priori unknown optimal steady states are a tradeoff between reaching the desired concentration $\overbar{c}_t$ and minimizing the coolant flow rate $u_t$. Even though the cost functions $L_t$ are convex, the resulting steady-state cost functions $L^s_t(\vartheta)$ are nonconvex due to the nonlinear system dynamics. More specifically, the steady-state cost functions $L^s_t(\vartheta)$ are only locally convex in a neighborhood of their respective minimum, compare Figure~\ref{fig:cost}.
In our simulations, the weighting~$q_t$ and desired concentration$~\overbar{c}_t$ are given by
\begin{align*}
	q_t &= 150-100\sin\left(\frac{2\pi\tau}{T} t\right) \\
	\overbar{c}_t &= 
	\begin{cases} 
		0.27+(0.65-0.27)\frac{\tau t}{90}  &\text{for $t\in[0,900$)} \\
		0.65 &\text{for $t\in[900,1800)$} \\
		0.65 - (0.65-0.3)\left(\frac{\tau t-180}{60}\right) &\text{for $t\in[1800,2400]$}
	\end{cases}
\end{align*}
where $T = 2400$ ($240$\,s) is the total length of the simulation.

\subsection{Design of the stabilizing controller}

We compute a stabilizing controller of the form $g(x,v)=K(v)\big(x-h(v)\big)$ using~\cite[Alg. 2]{Koehler20}. Additionally, we obtain a Lyapunov function  $V(x,v)=\norm{x-h(v)}^2_{P(v)}$ with $P(v)\succ0$ for all $v\in\intSv$ from this approach. Hence, Assumption~\ref{ass:stability} holds for this choice of stabilizing controller.

\subsection{Design of the reference governor} 

We proceed to design the RG and the safe set $\MAS$ in the OCO-RG framework. In this simulation, we compare two methods to design the safe set $\MAS$, both of which satisfy Assumption~\ref{ass:MAS}: 
\begin{enumerate}
	\item First, we compute $V_\mathrm{max}>0$ such that $V(x,v)\leq V_\mathrm{max}$ implies $\Phi(x,v,t)\in\constraintset$ for all $t\in\mathbb{N}$. We obtain $V_\mathrm{max}=0.0135$, and let $\MAS_\mathrm{fix}=\{(x,v)~|~V(x,v)\leq V_\mathrm{max}\}$. 
	\item Second, we compute a continuous function $\Gamma(v)$ such that $V(x,v)\leq\Gamma(v)$ implies $\Phi(x,v,t)\in\constraintset$ for all $t\in\mathbb{N}$, and set $\MAS_\mathrm{var}=\{(x,v)~|~V(x,v)\leq \Gamma(v)\}$, compare~\cite{Garone16} and Approach~2 in Appendix~\ref{appendix:MAS} for more details. 
\end{enumerate}
The resulting safe sets $\MAS_\mathrm{fix}$ and $\MAS_\mathrm{var}$ are illustrated in Figure~\ref{fig:steady-states}. Since $v$, i.e., the temperature $\vartheta$, is scalar, we apply the scalar RG~\eqref{eq:scalar_RG}, which can be solved efficiently online via bisection and satisfies Assumption~\ref{ass:general_RG}. 

\subsection{Design of the OCO algorithm}

We compare two different OCO algorithms within the proposed OCO-RG framework in our simulations: 
\begin{enumerate}
	\item OGD \cite{Zinkevich03}: Online gradient descent with step size $\gamma=2.5\cdot10^{-4}$, i.e., 
	\[
		r_t= \Pi_{\intSv} \big( r_{t-1} - \gamma \nabla L^s_t(r_{t-1}) \big).
	\]
	\item Prev. opt.: An algorithm that solves the steady-state optimization problem to optimality and sets $r_t=\eta_{t-1}$.
\end{enumerate}
For the latter, we run gradient descent with the same step size as the OGD algorithm until the norm of the gradient is smaller than the tolerance $10^{-9}$. The algorithm prev. opt. directly satisfies $Q$-linear convergence~\eqref{eq:exp_convergence} with $\kappa=0$. For OGD, we verified numerically that $Q$-linear convergence~\eqref{eq:exp_convergence} holds with $\kappa\approx0.986$ for all $q_t\in[50,250]$ and $\overbar{c}_t\in[0.25,0.65]$. Hence, the assumptions of Corollary~\ref{cor:regret_exp_conv_OCO_algo} are satisfied, and, thus, the regret bound~\eqref{eq:cor_bound} holds for all the considered algorithms.

\subsection{Results}

\begin{figure}
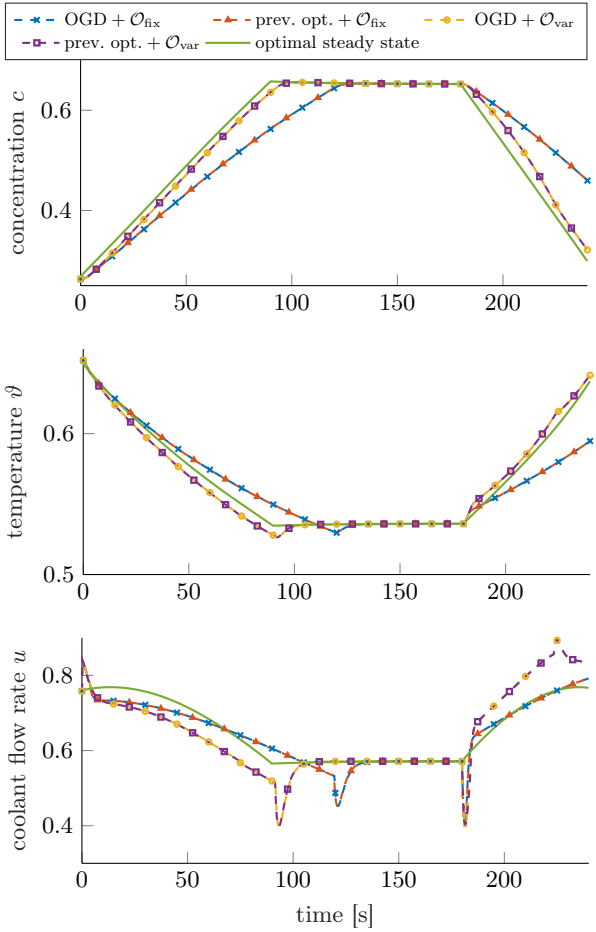
	
	\centering
	\small
	\setlength\hohe{3cm}
	\setlength\breite{.4\textwidth}
	\include{figures/concentration}
	\vspace{-8ex}
	\include{figures/temperature}
	\vspace{-8ex}
	\include{figures/flow}
	\vspace{-8ex}
	\caption{Closed-loop trajectories for the four combinations of OCO algorithms and RGs, together with the optimal steady states. From top to bottom: Concentration $c_t$, temperature $\vartheta_t$, control input $u_t$.}
	\label{fig:closed-loop}	
\end{figure}

The offline computations to obtain the stabilizing controller $g$, the Lyapunov function $V$, and the safe sets $\MAS_\mathrm{fix}$ and $\MAS_\mathrm{var}$ (i.e., $V_\mathrm{max}$ and $\Gamma(v)$) were completed\footnote{All experiments were performed on a standard laptop (Intel Core i9 with $2.6$\,GHz and $16$\,GB RAM under Windows 10) in MATLAB.} in less than $4$\,s. The resulting closed-loop concentration $c_t$, temperature $\vartheta_t$, and control input $u_t$ are shown in Figure~\ref{fig:closed-loop}. The normalized dynamic regret and the average computation time together with its standard deviation for each combination are reported in Table~\ref{tab:sim_results}. All combinations of RGs and OCO algorithms are able to track the optimal steady states. As expected, using the larger safe set $\MAS_\mathrm{var}$ yields faster convergence and lower regret, but requires more computation time due to repeated evaluations of $\Gamma(v)$ and $K(v)$ at each time instance. However, the computation times are very fast for each combination. Comparing the two OCO algorithms, the closed-loop trajectories are very close, and the added benefit of full optimization is negligible in this example. Full optimization requires considerably more computation time compared to OGD, but the influence of both OCO algorithms on the computational complexity is small compared to the RGs. To summarize, this example shows the flexibility of the proposed OCO-RG framework, which allows one to combine different RGs and OCO schemes trading off performance and computational efficiency.

\begin{table}
	\begin{center}
		\caption{Normalized dynamic regret and average computation times for evaluating the OCO algorithms and the RGs.} \label{tab:sim_results}
		\begin{tabular}{llrr@{$\pm$}lr@{$\pm$}l}
			&& \multirow{2}{1cm}{Norm. regret} & \multicolumn{4}{c}{Avg. comp. time [$\mathrm{\mu}$s]}\\
			&&& \multicolumn{2}{c}{OCO} & \multicolumn{2}{c}{RG} \\
			\hline
			OGD & $\MAS_\mathrm{fix}$ & $100.00\%$ & $0.057$&$\phantom{1}0.002$ & $46$&$\phantom{2}27$ \\
			OGD & $\MAS_\mathrm{var}$ & $9.85\%$ & $0.057$&$\phantom{1}0.002$ & $189$&$275$ \\
			Prev. opt. & $\MAS_\mathrm{fix}$ & $99.92\%$ & $6.5\phantom{00}$&$10.1$ & $46$&$\phantom{2}26$ \\
			Prev. opt. & $\MAS_\mathrm{var}$ & $9.85\%$ & $6.5\phantom{00}$&$10.2$ & $184$&$273$
		\end{tabular}
	\end{center}
\end{table}

\section{Conclusion} \label{sec:conclusion}

In this work, we proposed a general framework for controlling nonlinear dynamical systems subject to time-varying and a priori unknown cost functions, and state and input constraints. This framework combines an online convex optimization algorithm that tracks the time-varying optimal setpoint, and a reference governor that ensures constraint satisfaction. In our analysis, we established an upper bound on the dynamic regret of the proposed framework, which depends linearly on both, the dynamic regret and the path length of the online convex optimization algorithm. For $Q$-linear convergent optimization algorithms, the dynamic regret of the proposed OCO-RG framework is bounded linearly only in the variation of the cost functions, i.e., the path length of the optimal steady-state references. Implementation of the proposed approach is showcased on a numerical experiment of a nonlinear chemical reactor, where we demonstrate modularity and flexibility of our framework.

The main limitation of our work is the assumption of exact knowledge of the dynamical system and no uncertainty. Therefore, future work includes robustifying the proposed OCO-RG scheme with respect to these uncertainties. Furthermore, deriving an improved (sublinear) regret bound with respect to the OCO algorithm's path length and studying corresponding lower bounds is an interesting avenue for future research.

\appendix
\section*{Appendices}
\section{Proof of Proposition~\ref{prop:bounded_regret_path_length_exp_convergence}} \label{appendix:Prop_bounded_regret_path_length_exp_convergence}
\begin{pf}
	First, using the triangle inequality and \eqref{eq:exp_convergence} we obtain
	\begin{align*}
		&\sum_{t=1}^T \norm{r_t-\eta_t}_S \leq \sum_{t=1}^T \norm{r_t - \eta_{t-1}}_S + \sum_{t=1}^T \norm{\eta_t-\eta_{t-1}}_S \\
		&\refleq{\eqref{eq:exp_convergence}} \kappa \sum_{t=1}^T \norm{r_{t-1}-\eta_{t-1}}_S + \sum_{t=1}^T \norm{\eta_t-\eta_{t-1}}_S \\
		&\mkern8mu\leq \kappa \norm{r_0-\eta_0}_S + \kappa \sum_{t=1}^T \norm{r_{t}-\eta_{t}}_S + \sum_{t=1}^T \norm{\eta_t-\eta_{t-1}}_S.
	\end{align*}
	From here, rearranging yields
	\begin{align}
		&\sum_{t=1}^T \norm{r_t-\eta_t}_S \leq c_\kappa\norm{r_0-\eta_0}_S + c_\kappa \sum_{t=1}^T \norm{\eta_t-\eta_{t-1}}_S \nonumber \\
		&\leq c_\kappa\norm{S^\frac{1}{2}}\norm{r_0-\eta_0} + c_\kappa\norm{S^\frac{1}{2}} \sum_{t=1}^T \norm{\eta_t-\eta_{t-1}} \label{eq:prop_1}
	\end{align}
	where $c_\kappa:=\frac{\kappa}{1-\kappa}$. Note that the steady-state cost functions $\ststcost(v)$ are $l_s$-Lipschitz continuous on $\stst_\mathrm{v}$, compare Appendix~\ref{appendix:Lipschitz}. Furthermore, $r_t,\eta_t\in\intSv\subseteq\stst_\mathrm{v}$ for all $t\in\mathbb{N}$. Thus, it follows that
	\begin{align*}
		&\regret^\mathrm{OCO}=\sum_{t=0}^T \ststcost(r_t) - \ststcost(\eta_t) \leq l_s \sum_{t=0}^T \norm{r_t-\eta_t} \\
		\leq \mkern9mu& l_s \norm{S^{-\frac{1}{2}}} \sum_{t=0}^T \norm{r_t-\eta_t}_S \\
		\refleq{\eqref{eq:prop_1}} \,& \tilde{c}_\mathrm{OCO,0} \norm{r_0-\eta_0} + \tilde{c}_\mathrm{OCO} \sum_{t=1}^T \norm{\eta_t-\eta_{t-1}},
	\end{align*}
	where we define 
	\begin{align}
		\tilde{c}_\mathrm{OCO,0} &:= l_s\norm{S^{-\frac{1}{2}}}\left(1+c_\kappa\norm{S^{\frac{1}{2}}}\right),\label{eq:cOCO0}\\
		\tilde{c}_\mathrm{OCO} &:= l_s\norm{S^{-\frac{1}{2}}}c_\kappa\norm{S^{\frac{1}{2}}}. \label{eq:cOCO}
	\end{align}
	Moreover, \eqref{eq:prop_1} together with \eqref{eq:exp_convergence} implies
	\begin{align*}
		&\sum_{t=1}^T \norm{r_t-r_{t-1}}_S \leq \sum_{t=1}^T \norm{r_t-\eta_{t-1}}_S + \sum_{t=0}^{T-1}\norm{r_{t}-\eta_{t}}_S \\
		&\refleq{\eqref{eq:exp_convergence}} (1+\kappa) \norm{r_0-\eta_0}_S + (1+\kappa) \sum_{t=1}^T \norm{r_{t}-\eta_{t}}_S
	\end{align*}
	\begin{align*}
		&\refleq{\eqref{eq:prop_1}} (1+\kappa)\left( 1+c_\kappa\norm{S^{\frac{1}{2}}}\right) \norm{r_0-\eta_0} \\
		&\qquad + (1+\kappa)c_\kappa\norm{S^{\frac{1}{2}}} \sum_{t=1}^T \norm{\eta_t - \eta_{t-1}},
	\end{align*}
	Noting that $\regret^\mathrm{PL}\leq\norm{S^{-\frac{1}{2}}}\sum_{t=1}^T \norm{r_t-r_{t-1}}_S$ concludes the proof.\hfill$\square$
\end{pf}

\section{Proof of Lemma~\ref{lem:Lyap_fcn}} \label{appendix:Lemma_Lyap_fcn}
\begin{pf}
	This proof follows established methods for converse Lyapunov theorems in the literature (compare, e.g., \cite[Theorem~1]{Jiang02}), but we additionally require~\eqref{eq:Lipschitz_Lyap_fcn}. We construct a function $V(x,v)$ that satisfies the desired properties \eqref{eq:Lyap_fcn_bounds}-\eqref{eq:Lipschitz_Lyap_fcn}. To this end, let
	\[
	V(x,v) := \sum_{i=0}^{N-1} \norm{\Phi(x,v,i)-h(v)}, 
	\]
	where $N\in\mathbb{N}$ satisfies $c_\Phi\lambda^N < 1$ with $c_\Phi\geq1$ and $\lambda\in[0,1)$ from Assumption~\ref{ass:stability}. It follows that
	\begin{align*}
		&\norm{x-h(v)} \leq V(x,v) = \sum_{i=0}^{N-1} \norm{\Phi(x,v,i)-h(v)} \\
		\refleq{\eqref{eq:stability_bound}}\, &c_\Phi \norm{x-h(v)} \sum_{i=0}^{N-1} \lambda^i \leq \frac{c_\Phi}{1-\lambda}\norm{x-h(v)},
	\end{align*}
	i.e., \eqref{eq:Lyap_fcn_bounds} holds with $\lambda_1 = 1$ and $\lambda_2 = \frac{c_\Phi}{1-\lambda}$. Furthermore,
	\begin{align*}
		&V\big(f_g(x,v),v\big)-V(x,v) \\
		= \,& \norm{\Phi\big(f_g(x,v),v,N-1\big)-h(v)} - \norm{x-h(v)} \\
		= \,& \norm{\Phi(x,v,N)-h(v)} - \norm{x-h(v)} \\
		\refleq{\eqref{eq:stability_bound}} \,&-(1-c_\Phi\lambda^N) \norm{x-h(v)},
	\end{align*}
	i.e., \eqref{eq:Lyap_fcn_decrease} holds with $\lambda_3 = 1-c_\Phi\lambda^N > 0$ since $c_\Phi\lambda^N<1$. 
	
	To prove~\eqref{eq:Lipschitz_Lyap_fcn}, let $x\in\mathcal{X}$ and $v,\tilde{v}\in\stst_\mathrm{v}$ be such that $\Phi(x,v,t)\in\mathcal{X}$ and $\Phi(x,\tilde{v},t)\in\mathcal{X}$ hold for all $t\in\mathbb{N}$. Then, the reverse triangle inequality yields
	\begin{align*}
		&\norm{V(x,v) - V(x,\tilde{v})} \\
		= \, &\norm{ \sum_{i=0}^{N-1} \Big( \norm{ \Phi(x,v,i) - h(v)} - \norm{\Phi(x,\tilde{v},i) - h(\tilde{v})} \Big) }  \\
		\leq \,&\sum_{i=0}^{N-1} \norm{ \Phi(x,v,i) - \Phi(x,\tilde{v},i) + h(\tilde{v}) - h(v)} \\
		\leq \,&\sum_{i=0}^{N-1} \Big( \norm{h(v) - h(\tilde{v})} + \norm{ \Phi(x,v,i) - \Phi(x,\tilde{v},i)} \Big).
	\end{align*}
	Note that $\Phi(x,v,i)\in\mathcal{X}$ and $\Phi(x,\tilde{v},i) \in\mathcal{X}$ hold for any $i\in\mathbb{N}_{[0,N-1]}$ by assumption. Combining this with Assumption~\ref{ass:Lipschitz_system}, we get
	\begin{align*}
		&\norm{ \Phi(x,v,i) - \Phi(x,\tilde{v},i)} \\
		=\,& \norm{ f_g\big(\Phi(x,v,i-1),v\big) - f_g\big(\Phi(x,\tilde{v},i-1),\tilde{v}\big) } \\
		\leq \,&l_f \norm{ \Phi(x,v,i-1) - \Phi(x,\tilde{v},i-1)} + l_f \norm{v-\tilde{v}} \\ 
		\leq \,& \sum_{k=1}^{i} l_f^k \norm{v-\tilde{v}}.
	\end{align*}
	Plugging this in and using Assumption~\ref{ass:Lipschitz_system}, we obtain
	\begin{align*}
		&\norm{V(x,v) - V(x,\tilde{v})} \\
		\leq \,& \sum_{i=0}^{N-1} \norm{h(v) - h(\tilde{v})} + \sum_{i=1}^{N-1} \sum_{k=1}^{i} l_f^k \norm{v-\tilde{v}} \\
		\leq \,& Nl_h \norm{v - \tilde{v}} + \norm{v-\tilde{v}} (N-1) \sum_{k=1}^{N-1} l_f^k,
	\end{align*}
	i.e., \eqref{eq:Lipschitz_Lyap_fcn} with $l_V = Nl_h + (N-1) \sum_{k=1}^{N-1} l_f^k$. \hfill$\square$
\end{pf}

\section{Proof of Lemma~\ref{lem:Lyap_fcn_increase_bound}} \label{appendix:Lemma_Lyap_fcn_increase_bound}

\begin{pf}
	For any $\tau_1,\tau_2\in\mathbb{N}$ such that $\tau_2\geq\tau_1$, we have $(x_t,v_t)\in\MAS$ for all $t\in\mathbb{N}_{[\tau_1,\tau_2]}$ by Assumption~\ref{ass:general_RG}. Thus, $\Phi(x_{\tau_i},v_{\tau_i},t)\in\mathcal{X}$, $i\in\{1,2\}$, for all $t\in\mathbb{N}$ by Assumption~\ref{ass:MAS}, and we can apply Lemma~\ref{lem:Lyap_fcn} to get
	\begin{align*}
		&V(x_{\tau_2},v_{\tau_2}) \\
		\leq \,& V(x_{\tau_2},v_{\tau_2-1}) + \norm{V(x_{\tau_2},v_{\tau_2})-V(x_{\tau_2},v_{\tau_2-1})} \\
		\refleq{\eqref{eq:Lipschitz_Lyap_fcn}} &V\big(f_g(x_{\tau_2-1},v_{\tau_2-1}),v_{\tau_2-1}\big) + l_V \norm{v_{\tau_2}-v_{\tau_2-1}} \\
		\refleq{\eqref{eq:Lyap_fcn_decrease}} &V(x_{\tau_2-1},v_{\tau_2-1}) - \lambda_3\norm{x_{\tau_2-1}-h(v_{\tau_2-1})} \\
		&\qquad + l_V \norm{v_{\tau_2}-v_{\tau_2-1}} \\
		\refleq{\eqref{eq:Lyap_fcn_bounds}} &\tilde{\lambda} V(x_{\tau_2-1},v_{\tau_2-1}) + l_V \norm{v_{\tau_2}-v_{\tau_2-1}}.
	\end{align*}
	Repeatedly applying these arguments yields the desired result~\eqref{eq:Lyap_recursion}. It remains to show the upper bound $V(x_t,v_t)\leq\overbar{V}$. For this, let $\diamS := \max_{u,v\in\intSv} \norm{u-v}$, which is bounded because $\intSv$ is compact. Then, for any $t\in\mathbb{N}$, using \eqref{eq:Lyap_recursion} with $\tau_2=t$ and $\tau_1=0$ yields
	\begin{align*}
		V(x_{t},v_{t}) &\leq \tilde{\lambda}^{t} V(x_{0},v_{0}) + l_V\sum_{i=1}^{t} \norm{v_i-v_{i-1}} \tilde{\lambda}^{t-i} \\
		&\!\refleq{\eqref{eq:Lyap_fcn_bounds}} \lambda_2 \norm{x_0-h(v_0)} + l_V \diamS \sum_{i=0}^{t-1} \tilde{\lambda}^i \\
		&\leq \overbar{\lambda}+\frac{l_V\diamS}{1-\tilde{\lambda}} =: \overbar{V},
	\end{align*}
	where $\overbar{\lambda}$ satisfies $\overbar{\lambda}\geq\lambda_2\norm{x-h(v)}$ for all $x\in\mathcal X_0$ and $v\in\intSv$, which exists because $\mathcal X_0$ and $\intSv$ are compact.\hfill$\square$
\end{pf}

\section{Proof of Lemma~\ref{lem:RG_avg_contr}} \label{appendix:Lemma_RG_avg_contr}
\begin{pf}
	First, define the level sets of $V(x,u)$ by $\mathcal V_\mu(v) := \{x\in\mathcal{X}~|~V(x,v)\leq\mu\}$ with some $\mu>0$. Due to~\eqref{eq:Lyap_fcn_bounds}, there exists $\mu>0$ such that $\mathcal V_\mu(v)\subseteq\ball[n]_\delta\big(h(v)\big)$ for any $v\in\intSv$ with $\delta>0$ from Assumption~\ref{ass:MAS}. Then, let $M$ be such that $\tilde{\lambda}^M \overbar V\leq\frac{\lambda_1\mu}{4\lambda_2}$, with $\overbar{V}$ and $\tilde{\lambda}$ from Lemma~\ref{lem:Lyap_fcn_increase_bound}, and $\lambda_1,\lambda_2,\lambda_3$ from Lemma~\ref{lem:Lyap_fcn}. We prove the desired result by constructing $\epsilon>0$ such that \eqref{eq:RG_avg_contr} is satisfied. To do so, fix any $t\in\mathbb{N}$. Note that, if $r_k = v_k$ for any $k\in\mathbb N_{[t,t+M]}$, the result is trivially true, because $1-\rho(\alpha_t)\in[0,1]$ by definition of $\rho$ implies
	\[
	\prod_{i=t}^{t+M} (1-\rho(\alpha_i)) \leq 1-\rho(\alpha_k) = 1-\rho(\rho^{-1}(\epsilon)) = 1-\epsilon.
	\]
	Therefore, in the following we only treat the case $r_k\neq v_k$ for all $k\in\mathbb N_{[t,t+M]}$. We proceed by a case distinction.
	
	\underline{Case 1:} $\sum_{i=t}^{t+M-1} \alpha_i > \frac{\lambda_1\mu}{4\lambda_2l_V}$, where $l_V,\lambda_1,\lambda_2$ are from Lemma~\ref{lem:Lyap_fcn}. In this case, there exists $i^*\in\mathbb{N}_{[t,t+M]}$ such that $\alpha_{i^*}>\frac{\lambda_1\mu}{4\lambda_2l_VM}$ holds, which implies $1-\rho(\alpha_{i^*})<1-\rho(\frac{\lambda_1\mu}{4\lambda_2l_VM})$. Thus, we get
	\[
	\prod_{i=t}^{t+M}(1-\rho(\alpha_i)) \leq 1-\rho(\alpha_{i^*}) < 1-\rho\left(\frac{\lambda_1\mu}{4\lambda_2l_VM}\right).
	\]
	
	\underline{Case 2:} $\sum_{i=t}^{t+M-1} \alpha_i \leq \frac{\lambda_1\mu}{4\lambda_2l_V}$. 
	Recalling that $V(x_t,v_t)\leq\overbar V$ holds for all $t\in\mathbb{N}$, $\tilde{\lambda}^M \overbar{V}\leq \frac{\lambda_1\mu}{4\lambda_2}$, and using \eqref{eq:Lyap_recursion} with $\tau_2=t+M-1$ and $\tau_1=t$, we obtain
	\begin{align}
		&V(x_{t+M},v_{t+M-1}) = V\big(f_g(x_{t+M-1},v_{t+M-1}),v_{t+M-1}\big) \nonumber \\
		&\refleq{\eqref{eq:Lyap_fcn_bounds},\eqref{eq:Lyap_fcn_decrease}} \tilde{\lambda} V(x_{t+M-1},v_{t+M-1}) \nonumber \\
		&\mkern9mu\refleq{\eqref{eq:Lyap_recursion}} \tilde{\lambda}^M V(x_{t},v_{t}) + l_V \sum_{i=t+1}^{t+M-1} \norm{v_i-v_{i-1}} \tilde{\lambda}^{t+M-i-1} \nonumber \\
		&\mkern13mu\leq \tilde{\lambda}^M \overbar V + l_V \sum_{i=t+1}^{t+M-1} \norm{v_i-v_{i-1}} \nonumber \\
		&\mkern9mu\refeq{\eqref{eq:def_alpha}} \tilde{\lambda}^M \overbar V + l_V \sum_{i=t+1}^{t+M-1} \alpha_i 
		\leq 
		\frac{\lambda_1\mu}{2\lambda_2}. \label{eq:aux_bound_1}
	\end{align}
	Hence, for any $\nu\in\intSv$ such that $\norm{\nu-v_{t+M-1}}\leq\frac{\mu}{2\lambda_2l_h}$, we obtain
	\begin{align*}
		&V(x_{t+M},\nu) \refleq{\eqref{eq:Lyap_fcn_bounds}} \lambda_2\norm{x_{t+M}-h(\nu)} \\
		\leq \mkern11mu&\lambda_2 \norm{x_{t+M}-h(v_{t+M-1})} + \lambda_2 \norm{h(\nu)-h(v_{t+M-1})} \\
		\refleq{\eqref{eq:Lyap_fcn_bounds}} \mkern7mu &\frac{\lambda_2}{\lambda_1} V(x_{t+M},v_{t+M-1}) + \lambda_2 l_h \norm{\nu-v_{t+M-1}} \\
		\refleq{\eqref{eq:aux_bound_1}} \, &\frac{\mu}{2}+\frac{\mu}{2} = \mu,
	\end{align*}
	i.e., $x_{t+M}\in\mathcal{V}_\mu(\nu)\subseteq\ball[n]_\delta\big(h(\nu)\big) \subseteq \MAS_\mathrm{x}(\nu)$ by Assumption~\ref{ass:MAS}. Since $\nu\in\intSv$ was arbitrary, this implies $\{v\in\intSv~|~\norm{v-v_{t+M-1}}\leq\frac{\mu}{2\lambda_2l_h}\}\subseteq\MAS_\mathrm{v}(x_{t+M})$. Therefore, the constant $\overbar{\nu}_{t+M}$ in Assumption~\ref{ass:general_RG} satisfies $\overbar{\nu}_{t+M}\geq\frac{\mu}{2\lambda_2l_h}$. Recalling $v_{t+M}\neq r_{t+M}$, we get $\alpha_{t+M} = \norm{v_{t+M}-v_{t+M-1}} \geq \overbar{\nu}_{t+M} \geq \frac{\mu}{2\lambda_2l_h}$ by Assumption~\ref{ass:general_RG}. Hence, we have
	\begin{equation*}
		\prod_{i=t}^{t+M} (1-\rho(\alpha_i)) \leq 1-\rho\left(\alpha_{t+M}\right) \leq 1-\rho\left(\frac{\mu}{2\lambda_2l_h}\right).
	\end{equation*}
	Finally, combining the two cases, we get the result with $\epsilon = \min\left( \rho(\frac{\lambda_1\mu}{4\lambda_2l_VM}),\rho(\frac{\mu}{2\lambda_2 l_h})\right) \in(0,1]$.\hfill$\square$
\end{pf}

\section{Proof of Theorem~\ref{thm}} \label{appendix:thm}
\begin{pf}	
	Constraint satisfaction $(x_t,u_t)\in\constraintset$ for all $t\in\mathbb{N}$ directly follows from Assumptions~\ref{ass:MAS} and~\ref{ass:general_RG}. Furthermore, we note that the steady-state cost functions $L^s_t$ are $l_s$-Lipschitz continuous, as shown in Appendix~\ref{appendix:Lipschitz}. Using Assumptions~\ref{ass:cost_fcn} and \ref{ass:Lipschitz_system} yields
	\begin{align}
		\regret &= \sum_{t=0}^T L_t(x_t,u_t) - \ststcost(\eta_t) \nonumber \\
		&= \sum_{t=0}^T L_t\big(x_t,g(x_t,v_t)\big) - \ststcost(v_t) \nonumber \\ 
		&\qquad+ \sum_{t=0}^T \ststcost(v_t) - \ststcost(r_t) + \sum_{t=0}^T \ststcost(r_t) - \ststcost(\eta_t) \nonumber \\
		\begin{split} \label{eq:regret_bound_1}
			&\mkern-18mu\refleq{\eqref{eq:Lipschitz_cost},\eqref{eq:OCO-regret}} l(1+l_g) \sum_{t=0}^T \norm{x_t-h(v_t)} \\
			&\qquad+ l_s \sum_{t=0}^T \norm{v_t - r_t} + \regret^{OCO}.
		\end{split}
	\end{align}
	We bound the two sums in \eqref{eq:regret_bound_1} separately. First, recall that $r_t\notin\MAS_\mathrm{v}(x_t)$ implies $r_t\neq v_t$ by Assumption~\ref{ass:general_RG}, and, thus, $\alpha_t=\norm{v_t-v_{t-1}}$ by the definition of $\alpha_t$ in \eqref{eq:def_alpha}. Furthermore, if $r_t\in\MAS_\mathrm{v}(x_t)$, then $v_t=r_t$ again by Assumption~\ref{ass:general_RG}. Therefore, $\norm{v_t-r_t}\leq(1-\rho(\alpha_t))\norm{r_t-v_{t-1}}$ holds for all $t\in\mathbb{N}$. Hence, we obtain
	\begin{align*}
		&\norm{v_t-r_t} \leq \big(1-\rho(\alpha_t)\big) \norm{v_{t-1}-r_t} \\
		\leq \,& \big(1-\rho(\alpha_t)\big) \norm{v_{t-1}-r_{t-1}} + \big(1-\rho(\alpha_t)\big) \norm{r_t - r_{t-1}}.
	\end{align*}
	Applying the same arguments repeatedly yields
	\begin{align*}
		\norm{v_t-r_t} \leq \,&\left( \prod_{i=1}^t \big(1-\rho(\alpha_i)\big) \right) \norm{r_0-v_0} \\
		&\qquad+ \sum_{k=1}^t \left(\prod_{i=k}^{t} \big(1-\rho(\alpha_i)\big) \right) \norm{r_k - r_{k-1}}.
	\end{align*}
	Then, summing over $t$, noting $r_0=v_0$, and applying Lemma~\ref{lem:RG_avg_contr}, we obtain
	\begin{align}
		&\sum_{t=0}^T\norm{v_t-r_t}\leq \sum_{t=0}^T \sum_{k=1}^t \!\left(\!\left(\prod_{i=k}^{t} \big(1-\rho(\alpha_i)\big) \!\right)\! \norm{r_k - r_{k-1}}\!\right) \nonumber \\
		&\mkern9mu\leq \sum_{t=1}^T \left( \norm{r_t - r_{t-1}} \sum_{k=t}^\infty \left(\prod_{i=t}^{k} \big(1-\rho(\alpha_i)\big) \right) \right) \nonumber \\
		&\refleq{\eqref{eq:RG_avg_contr}} \sum_{t=1}^T \left( \norm{r_t - r_{t-1}} \sum_{j=0}^\infty \Big(M(1-\epsilon)^j \Big) \right) \nonumber \\
		&\mkern9mu\leq \frac{M}{\epsilon} \sum_{t=1}^T  \norm{r_t - r_{t-1}}
		\refeq{\eqref{eq:path-length-regret}} \frac{M}{\epsilon} \regret^{PL}. \label{eq:vt-rt}
	\end{align}		
	It remains to bound the first sum in \eqref{eq:regret_bound_1}. To do so, we first derive a bound on the path length of the reference input. Recalling $v_0=r_0$, we get
	\begin{align}
		&\sum_{t=1}^T \norm{v_t-v_{t-1}} \leq \sum_{t=1}^T \norm{v_t - r_{t-1}} + \sum_{t=1}^T \norm{v_{t-1}-r_{t-1}} \nonumber \\
		&\refleq{\eqref{eq:path-length-regret}} 2\sum_{t=1}^T \norm{v_t - r_t} + \regret^{PL}
		\refleq{\eqref{eq:path-length-regret},\eqref{eq:vt-rt}} c_\epsilon \regret^{PL}, \label{eq:vt-vt-1}
	\end{align}
	where $c_\epsilon:=\frac{2M+\epsilon}{\epsilon}$. Next, we bound the first sum in \eqref{eq:regret_bound_1}. Using Lemma~\ref{lem:Lyap_fcn} and \eqref{eq:Lyap_recursion} with $\tau_2=t$ and $\tau_1=0$ yields
	\begin{align*}
		&\norm{x_t-h(v_t)} \refleq{\eqref{eq:Lyap_fcn_bounds}} \frac{1}{\lambda_1} V(x_t,v_t) \\
		\refleq{\eqref{eq:Lyap_recursion}} &\frac{1}{\lambda_1} \tilde{\lambda}^{t} V(x_0,v_0) + \frac{l_V}{\lambda_1} \sum_{i=1}^{t} \norm{v_i-v_{i-1}} \tilde{\lambda}^{t-i} \\
		\refleq{\eqref{eq:Lyap_fcn_bounds}}\mkern7mu &\frac{\lambda_2}{\lambda_1} \tilde{\lambda}^{t} \norm{x_0-h(v_0)} + \frac{l_V}{\lambda_1} \sum_{i=1}^{t} \norm{v_i-v_{i-1}} \tilde{\lambda}^{t-i}.
	\end{align*}
	Finally, summing the above inequality over $t$ yields
	\begin{align*}
		&\sum_{t=0}^T \norm{x_t-h(v_t)} \leq \frac{\lambda_2}{\lambda_1} \norm{x_0-h(v_0)} \sum_{t=0}^T \tilde{\lambda}^t \\
		&\hspace{.17\textwidth}+ \frac{l_V}{\lambda_1} \sum_{t=0}^T \sum_{i=1}^{t} \norm{v_{i} - v_{i-1}} \tilde{\lambda}^{t-i} \\
		\leq\mkern6mu& c_\lambda\norm{x_0-h(v_0)} + \frac{l_V}{\lambda_1} \sum_{t=1}^T \left( \norm{v_{t}-v_{t-1}} \sum_{i=0}^T \tilde\lambda^i \right) \\
		\leq\mkern6mu & c_\lambda\norm{x_0-h(v_0)} + \frac{l_Vc_\lambda}{\lambda_2} \sum_{t=1}^T \norm{v_{t}-v_{t-1}} \\
		\refeq{\eqref{eq:vt-vt-1}} & c_\lambda \norm{x_0-h(\eta_0)} + c_\lambda l_h\norm{v_0-\eta_0} +\frac{l_Vc_\lambda c_\epsilon}{\lambda_2} \regret^{PL},
	\end{align*}
	where $c_\lambda := \frac{\lambda_2}{\lambda_1(1-\tilde{\lambda})}$.\hfill{$\square$}
\end{pf}

\section{Proof of Proposition~\ref{prop:optimal_lower_bound}} \label{appendix:Prop_optimal_lower_bound}
\begin{pf}
	By the definition of dynamic regret in~\eqref{eq:def_regret} and Definition~\ref{ass:OCO-guarantees}, respectively, we have
	\begin{align*}
		&\regret\big(x_0,u_0,\dots,u_T\big) = \sum_{t=0}^T L_t(x_t,u_t)-L^s_t(\eta_t) \\
		=\,&\regret^\mathrm{OCO}(r_0,\dots,r_T) + \sum_{t=0}^T L_t(x_t,u_t) - L^s_t(r_t).
	\end{align*}
	Thus, it remains to be shown that there exists a sequence of cost functions such that $\sum_{t=0}^T L_t(x_t,u_t)-L^s_t(r_t) \geq 0$ holds. To this end, note that the cost functions $L_t$ (and, hence, also $L^s_t$) are a priori unknown. More specifically, at each time $t\in\mathbb{N}$, they are only revealed after the algorithm $\mathcal{A}^\mathrm{OCO}$ applies $r_t=\mathcal{A}^\mathrm{OCO}\left(\mathcal{I}_t\right)$. Thus, the cost functions $L_t$ may be chosen adversarially, i.e., depending on $r_t$. Hence, choosing
	\[
	L_t(x,u) = \norm{ \begin{bmatrix} x-h(r_t) \\ u-g(h(r_t),r_t) \end{bmatrix} }^2
	\]
	satisfies $0 = L^s_t(r_t) \leq L_t(x_t,u_t)$ for all $t\in\mathbb{N}$. \hfill$\square$
\end{pf}

\section{Lipschitz continuity of $\ststcost$} \label{appendix:Lipschitz}

\begin{lemma} \label{lem:Lipschitz}
	Suppose Assumptions~\ref{ass:steady-state-map},~\ref{ass:Lipschitz_system} and~\ref{ass:cost_fcn} hold. The steady-state cost functions $\ststcost(v)=L_t\big(h(v),g(h(v),v)\big)$ are $l_s$-Lipschitz continuous on $\stst_\mathrm{v}$ for all $t\in\mathbb{N}$, where $l_s:=l(l_h+l_g+l_gl_h)$.
\end{lemma}
\begin{pf}
	For any $v_1,v_2\in\stst_\mathrm{v}$, we have $(h(v_1),v_1)\in\constraintset_g$ and $(h(v_2),v_2)\in\constraintset_g$, i.e., $h(v_1)\in\mathcal{X}$ and $h(v_2)\in\mathcal{X}$ hold. Thus, Assumptions~\ref{ass:Lipschitz_system} and~\ref{ass:cost_fcn} are applicable, and using Lipschitz continuity, we obtain
	\begin{align*}
		&\norm{\ststcost(v_1){-}\ststcost(v_2)} \\
		= \, &\norm{ L_t\big(h(v_1),g(h(v_1),v_1)\big) - L_t\big(h(v_2),g(h(v_2),v_2)\big) } \\
		\leq\, &l \big( \norm{h(v_1)-h(v_2)} + \norm{g(h(v_1),v_1)-g(h(v_2),v_2)} \big) \\
		\leq\,& l\big( (l_h+l_g)\norm{v_1-v_2}+l_g\norm{h(v_1)-h(v_2)} \big) \\
		\leq\,& l \left(l_h+l_g+l_gl_h\right) \norm{v_1-v_2},
	\end{align*}
	which is the desired result.\hfill$\square$
\end{pf}

\section{Determination of the safe set $\MAS$} \label{appendix:MAS}

In this section, we describe two approaches to construct a safe set $\MAS$ that satisfies Assumption~\ref{ass:MAS} based on availability of a Lyapunov function, compare~\cite{Garone17}. To this end, we assume that a Lyapunov function $V(x,v)$ for system~\eqref{eq:sys} is known. Given Assumptions~\ref{ass:steady-state-map}-\ref{ass:stability}, Lemma~\ref{lem:Lyap_fcn} provides such a Lyapunov function. Furthermore, for many practical problems, a quadratic Lyapunov function of the form $V(x,v)=\norm{x-h(v)}_{P(v)}^2$ can be constructed. 

\textit{Approach 1 \cite{Bemporad98}:} By standard uniform bounds on $V$~\eqref{eq:Lyap_fcn_bounds}, there exists $V_\mathrm{min}>0$ such that $V(x,v)\leq V_\mathrm{min}$ and $v\in\intSv$ imply $(x,v)\in\constraintset_g$. Furthermore, since (exponential) Lyapunov functions $V$ decrease along trajectories of the system~\eqref{eq:Lyap_fcn_decrease}, there exists $k^*\in\mathbb{N}$ such that $V\big(\Phi(x,v,k^*),v\big)\leq V_\mathrm{min}$ holds for all $x\in\mathcal{X}$ and $v\in\intSv$, compare Lemma~\ref{lem:Lyap_fcn_increase_bound}. Then, the safe $\MAS$ is given by
\[
\MAS = \big\{(x,v)\in\mathcal{X}\times\intSv~|~\big(\Phi(x,v,t),v\big)\in\constraintset_g~\forall t\in\mathbb{N}_{[0,k^*]}\big\}.
\]
Since sublevel sets of the Lyapunov function $V(x,v)$ are forward invariant, it follows that $(x,v)\in\MAS$ implies $V(\Phi(x,v,t),v)\leq V_\mathrm{min}$ for all $t\in\mathbb{N}_{\geq k^*}$. Thus, $(x,v)\in\MAS$ implies that $\big(\Phi(x,v,t),v\big)\in\constraintset_g$ holds for all $t\in\mathbb{N}$. Furthermore, there exists $\delta>0$ such that $V(x,v)\leq V_\mathrm{min}$ for all $x\in\ball[n]_\delta\big(h(v)\big)$ by~\eqref{eq:Lyap_fcn_bounds}. This implies $\ball[n]_\delta\big(h(v)\big)\subseteq\MAS_\mathrm{x}(v)$ for all $v\in\intSv$ by forward invariance of the sublevel sets of the Lyapunov function. Thus, Assumption~\ref{ass:MAS} is satisfied. Note that both $V_\mathrm{min}$ and $k^*$ can be computed offline \cite{Garone17}.

\textit{Approach 2 \cite{Garone16,Nicotra15}:} Suppose there exists a continuous function $\Gamma:\intSv\mapsto\mathbb{R}_{>0}$ such that $V(x,v)\leq\Gamma(v)$ implies $(x,v)\in\constraintset_g$ for all $v\in\intSv$. Then, the safe set $\MAS$ is given by
\[
\MAS = \big\{(x,v)\in\mathcal{X}\times\intSv~|~V(x,v)\leq\Gamma(v) \big\}.
\]
Again, by forward invariance of sublevel sets of the Lyapunov function $V(x,v)$, we get that $(x,v)\in\MAS$ implies $V\big(\Phi(x,v,t),v\big)\leq\Gamma(v)$ for all $t\in\mathbb{N}$, and, thus, $\big(\Phi(x,v,t),v\big)\in\constraintset_g$ for all $t\in\mathbb{N}$. Furthermore, note that we can always choose $\Gamma(v)\geq V_\mathrm{min}>0$. Combining this with standard boundedness properties of the Lyapunov function $V$~\eqref{eq:Lyap_fcn_bounds}, we have that there exist $\delta>0$ such that $V(x,v)\leq\Gamma(v)$ for all $x\in\ball[n]_\delta(h(v))$ and $v\in\intSv$. This implies $\ball[n]_\delta\big(h(v)\big)\in\MAS_\mathrm{x}(v)$ for all $v\in\intSv$, i.e., Assumption~\ref{ass:MAS} is satisfied. As mentioned above, $\Gamma(v)=V_\mathrm{min}$ is always a possible, albeit conservative, choice. Alternatively, $\Gamma(v)$ can be computed in closed form if (i) the constraints are polytopic $\constraintset_g = \big\{(x,v)~|~Z_\mathrm{x} x + Z_\mathrm{v} v \leq z \big\}$, and (ii) there exists $P(v)\succ0$ that satisfies $V(x,v)\geq \norm{x-h(v)}_{P(v)}^2$ \cite{Garone16}. In this case, we obtain the function $\Gamma(v)$ by letting $z_\mathrm{x,v}(v) := z - Z_\mathrm{x} h(v) - Z_\mathrm{v} v$, and computing
\[
	\Gamma_i(v) := \left( \frac{\left[z_\mathrm{x,v}(v)\right]_i}{\norm{\left[Z_\mathrm{x}\right]_i P(v)^{-\frac{1}{2}}}} \right)^2
\]
for each row $\big[ Z_\mathrm{x} \big]_i$ of $Z_x$ and each entry $\big[ z_\mathrm{x,v}(v) \big]_i$ of $z_\mathrm{x,v}(v)$. The desired function is then given by 
\[
	\Gamma(v) = \min_{i\in\mathbb{N}_{[1,n_z]}} \Gamma_i(v).
\]
To see this, note that $Z_\mathrm{x}x + Z_\mathrm{v}v \leq z$ holds if and only if $Z_\mathrm{x}\big(x-h(v)\big) \leq z_\mathrm{x,v}(v)$ is satisfied. Hence, $V(x,v)\leq\Gamma(v)$ implies $\norm{x-h(v)}^2_{P(v)}\leq\Gamma_i(v)$, and, thus,
\begin{align*}
	&\left[ Z_\mathrm{x} \right]_i \big(x-h(v)\big) \leq \norm{ \left[ Z_\mathrm{x} \right]_i P(v)^{-\frac{1}{2}} P(v)^{\frac{1}{2}} \big(x-h(v)\big) } \\
	&\leq \norm{ \left[ Z_\mathrm{x} \right]_i P(v)^{-\frac{1}{2}} } \norm{ x-h(v) }_{ P(v)} \leq \left[ z_\mathrm{x,v}(v) \right]_i
\end{align*}
holds. Finally, $\Phi(x,v,t)\in\constraintset_g$ holds for all $t\in\mathbb{N}$ and $(x,v)\in\constraintset_g$ that satisfy $V(x,v)\leq\Gamma(v)$ due to forward invariance of sublevel sets of the Lyapunov function.


\bibliographystyle{abbrv}        
\bibliography{bib}           
\end{document}

%% file: figures/cost.tex
%
%
\definecolor{mycolor1}{rgb}{0.00000,0.44700,0.74100}%
\definecolor{mycolor2}{rgb}{0.85000,0.32500,0.09800}%
\definecolor{mycolor3}{rgb}{0.92900,0.69400,0.12500}%
\definecolor{mycolor4}{rgb}{0.49400,0.18400,0.55600}%
\definecolor{mycolor5}{rgb}{0.46600,0.67400,0.18800}%
\begin{tikzpicture}

\begin{axis}[%
width=0.951\breite,
height=\hohe,
at={(0\breite,0\hohe)},
scale only axis,
xmin=0.4,
xmax=0.8,
xlabel style={font=\color{white!15!black}},
xlabel={temperature $\vartheta$},
ymin=0,
ymax=1,
ytick = {0.2,0.4,0.6,0.8},
ylabel style={font=\color{white!15!black}},
ylabel={normalized cost},
axis background/.style={fill=white},
axis x line*=bottom,
axis y line*=left,
legend style={legend columns = 3,at={(\breite,\hohe)}, anchor=south east, legend cell align=left, align=left, draw=white!15!black, nodes={scale=0.75, transform shape}}
]
\addplot [color=mycolor1]
  table[row sep=crcr]{%
0.4	1\\
0.41	0.972812427597844\\
0.42	0.943692072764151\\
0.43	0.909216225399235\\
0.44	0.868449142909439\\
0.45	0.821096583617572\\
0.46	0.767284386912088\\
0.47	0.707548672777373\\
0.48	0.64284129052025\\
0.49	0.574503239629553\\
0.5	0.504191808162663\\
0.51	0.433762833105191\\
0.52	0.365121412572174\\
0.53	0.300062807905015\\
0.54	0.240128287840749\\
0.55	0.186497430723956\\
0.56	0.139930057315524\\
0.57	0.100760303673935\\
0.58	0.0689356046647936\\
0.59	0.0440869679828826\\
0.6	0.02561476935612\\
0.61	0.0127758223203062\\
0.62	0.00476124790353088\\
0.63	0.000759138357119107\\
0.64	0\\
0.65	0.00178586001737824\\
0.66	0.00550561034897894\\
0.67	0.0106398277296565\\
0.68	0.0167582661882119\\
0.69	0.0235127760608496\\
0.7	0.030627802289705\\
0.71	0.0378900106337231\\
0.72	0.0451380667036895\\
0.73	0.052253180860733\\
0.74	0.059150732053461\\
0.75	0.0657730800729551\\
0.76	0.0720835476145759\\
0.77	0.0780614805413839\\
0.78	0.0836982595440156\\
0.79	0.0889941255626359\\
0.8	0.0939556851211845\\
};
\addlegendentry{$q=50, \overbar{c}=0.3$}

\addplot [color=mycolor2]
  table[row sep=crcr]{%
0.4	1\\
0.41	0.976821866395932\\
0.42	0.948919746449798\\
0.43	0.915128316898656\\
0.44	0.874901639908318\\
0.45	0.828015409487841\\
0.46	0.774586155932224\\
0.47	0.715118446871843\\
0.48	0.65052961611122\\
0.49	0.582132629281107\\
0.5	0.511569102339003\\
0.51	0.440695082301539\\
0.52	0.371432897456984\\
0.53	0.305610734041425\\
0.54	0.244814957937996\\
0.55	0.190277372784119\\
0.56	0.142811526931482\\
0.57	0.102801541816025\\
0.58	0.0702369892837625\\
0.59	0.0447806111715361\\
0.6	0.0258531632336684\\
0.61	0.0127209040688254\\
0.62	0.00457486207423462\\
0.63	0.000595445056962466\\
0.64	0\\
0.65	0.00207392851135127\\
0.66	0.00618775961858473\\
0.67	0.0118033412222112\\
0.68	0.0184723377454699\\
0.69	0.0258298224358024\\
0.7	0.0335851730705196\\
0.71	0.0415118817885629\\
0.72	0.0494373625736973\\
0.73	0.0572334196270364\\
0.74	0.0648077297734129\\
0.75	0.0720964785603252\\
0.76	0.0790581529730522\\
0.77	0.0856684136715591\\
0.78	0.0919159290562808\\
0.79	0.0977990386823444\\
0.8	0.103323114550858\\
};
\addlegendentry{$q=250, \overbar{c}=0.3$}

\addplot [color=mycolor3]
  table[row sep=crcr]{%
0.4	1\\
0.41	0.965209503877347\\
0.42	0.925275917972843\\
0.43	0.877629545716771\\
0.44	0.821549521034388\\
0.45	0.757011072882678\\
0.46	0.684603992675956\\
0.47	0.605568316696776\\
0.48	0.521801566525112\\
0.49	0.435794536139856\\
0.5	0.350482453252501\\
0.51	0.269019865658205\\
0.52	0.194507046408862\\
0.53	0.1297095165615\\
0.54	0.0768161162309985\\
0.55	0.0372733354764863\\
0.56	0.0117167620202983\\
0.57	0\\
0.58	0.00130344427294454\\
0.59	0.0142942880422501\\
0.6	0.037306487110054\\
0.61	0.0685136872192605\\
0.62	0.106076380594779\\
0.63	0.148253714727768\\
0.64	0.19347819662078\\
0.65	0.240396940812041\\
0.66	0.287885996189653\\
0.67	0.335045135681216\\
0.68	0.381179993567883\\
0.69	0.4257772324004\\
0.7	0.468476997904084\\
0.71	0.509045575592748\\
0.72	0.547350045063424\\
0.73	0.583335880905513\\
0.74	0.617007856413635\\
0.75	0.648414223843766\\
0.76	0.677633921808012\\
0.77	0.704766449613752\\
0.78	0.729924011606131\\
0.79	0.753225543048923\\
0.8	0.774792262447996\\
};
\addlegendentry{$q=150, \overbar{c}=0.5$}

\addplot [color=mycolor4]
  table[row sep=crcr]{%
0.4	0.320683088299799\\
0.41	0.299214016079886\\
0.42	0.279296877220974\\
0.43	0.256827598295541\\
0.44	0.231129860851693\\
0.45	0.202345526428317\\
0.46	0.171077671045365\\
0.47	0.138327263690164\\
0.48	0.105456448394849\\
0.49	0.0741193597814322\\
0.5	0.046146813283775\\
0.51	0.023391062153195\\
0.52	0.00755023652632046\\
0.53	0\\
0.54	0.00166067833696702\\
0.55	0.0129213375623158\\
0.56	0.0336301567684181\\
0.57	0.063146849311543\\
0.58	0.100441862232026\\
0.59	0.144221235803455\\
0.6	0.193055822504691\\
0.61	0.245497736200074\\
0.62	0.300173249589468\\
0.63	0.35584779880054\\
0.64	0.411463894800581\\
0.65	0.466156014294838\\
0.66	0.519248026835132\\
0.67	0.570238841833595\\
0.68	0.618781254294682\\
0.69	0.664657887805255\\
0.7	0.707756999082362\\
0.71	0.748049905281096\\
0.72	0.785571002191338\\
0.73	0.82040076845622\\
0.74	0.852651771313807\\
0.75	0.88245746212223\\
0.76	0.909963433249428\\
0.77	0.935320765696701\\
0.78	0.958681100945936\\
0.79	0.980193100635167\\
0.8	1\\
};
\addlegendentry{$q=50, \overbar{c}=0.65$}

\addplot [color=mycolor5]
  table[row sep=crcr]{%
0.4	0.328237074671675\\
0.41	0.311870050625493\\
0.42	0.293077147845134\\
0.43	0.270868902893579\\
0.44	0.245099334600608\\
0.45	0.216002258134384\\
0.46	0.184169268305505\\
0.47	0.150566646095921\\
0.48	0.116525993289748\\
0.49	0.0836879078789819\\
0.5	0.0538935949690514\\
0.51	0.0290323225175296\\
0.52	0.0108645577767631\\
0.53	0.00084862531205978\\
0.54	0\\
0.55	0.00880608975885477\\
0.56	0.0272074452342996\\
0.57	0.0546425690018764\\
0.58	0.0901419922111565\\
0.59	0.132450827115119\\
0.6	0.180158265169119\\
0.61	0.231816286835995\\
0.62	0.286036041652714\\
0.63	0.341556830389612\\
0.64	0.3972879120246\\
0.65	0.452326826404161\\
0.66	0.505959593250002\\
0.67	0.557648421351875\\
0.68	0.607011954449972\\
0.69	0.653802053258635\\
0.7	0.697879998315002\\
0.71	0.739193992613949\\
0.72	0.777759035132434\\
0.73	0.813639643632069\\
0.74	0.846935505431679\\
0.75	0.877769889678286\\
0.76	0.906280523229301\\
0.77	0.932612578649842\\
0.78	0.956913418575575\\
0.79	0.979328764940915\\
0.8	1\\
};
\addlegendentry{$q=250, \overbar{c}=0.65$}

\end{axis}

\begin{axis}[%
width=1.227\breite,
height=1.227\hohe,
at={(-0.16\breite,-0.135\hohe)},
scale only axis,
xmin=0,
xmax=1,
ymin=0,
ymax=.9,
axis line style={draw=none},
ticks=none,
axis x line*=bottom,
axis y line*=left
]
\end{axis}
\end{tikzpicture}